\documentclass[12pt]{article}
\usepackage[dvipdfmx]{graphicx}
\usepackage{amsmath}
\usepackage{amssymb}
\usepackage{subcaption}
\usepackage{authblk}
\usepackage{mathrsfs}
\usepackage{slashed}
\usepackage{color}
\usepackage{cite}
\usepackage{multirow}
\makeatletter
 
 \@addtoreset{equation}{section}
 \makeatother
\newcommand{\be}{\begin{equation}}
\newcommand{\ee}{\end{equation}}
\newcommand{\bea}{\begin{eqnarray}}
\newcommand{\eea}{\end{eqnarray}}
\newcommand{\beann}{\begin{eqnarray*}}
\newcommand{\eeann}{\end{eqnarray*}}
\newcommand{\nn}{\nonumber}
\newcommand{\ba}{\begin{array}}
\newcommand{\ea}{\end{array}}

\newcommand{\D}{\mathcal{D}}

\newcommand{\phib}{\bar{\phi}}

\newcommand{\psit}{\tilde{\psi}}

\newcommand{\Lt}{\tilde{L}}
\newcommand{\bphi}{\boldsymbol{\phi}}

\newcommand{\R}{{\mathbb R}}
\newcommand{\bse}{{\boldsymbol{e}}}

\newcommand{\mathscale}[2]{\text{\scalebox{#1}{${#2}$}}}

\DeclareMathOperator{\Ker}{Ker}

\DeclareMathOperator{\rank}{rank}
\DeclareMathOperator{\diag}{diag}

\DeclareMathOperator{\Tr}{Tr}

\DeclareMathOperator{\Pf}{Pf}

\title{Fermions and Zeta Function on the Graph}
\author[1]{So Matsuura\thanks{s.matsu@keio.jp}}
\author[2]{Kazutoshi Ohta\thanks{kazutoshi.ohta@mi.meijigakuin.ac.jp}}
\affil[1]{\it Research and Education Center for Natural Sciences,
Keio University, Yokohama, Kanagawa, Japan}
\affil[2]{\it Institute for Mathematical Informatics, Meiji Gakuin University, Yokohama, Kanagawa, Japan}

\date{}

\begin{document}
\maketitle

%\vspace*{0.5cm}

\begin{center}
{\bf Abstract}
\end{center}

We propose a novel fermionic model on the graphs. %way to construct fermions on graphs.
%We introduce a Dirac operator defined by a deformed incidence matrix on the graph, and 
%we show that the partition function of the fermion model is expressed in terms of the inverse of the graph zeta function.
The Dirac operator of the model consists of deformed incidence matrices on the graph %and mass terms 
and the partition function is given by the inverse of the graph zeta function.
We find that the coefficients of the inverse of the graph zeta function, 
which is a polynomial of finite degree in the coupling constant, 
%and we give an interpretation on the meaning of the polynomial from the viewpoint of fermionic cycles.
count the number of fermionic cycles on the graph. 
%We also give a construction of the fermion on grid graphs 
We also construct the model on grid graphs 
by using the concept of the covering graph and the Artin-Ihara $L$-function. 
In connection with this, we show that the fermion doubling is absent, and
the overlap fermions can be constructed on a general graph.
%Furthermore, we point out a relation between our model and statistical models by introducing the winding number around cycles, where the distribution of the poles of the graph zeta function plays a crucial role.
Furthermore, we relate our model to statistical models by introducing the winding number around cycles, where the distribution of the poles of the graph zeta function (the zeros of the partition function) plays a crucial role.
%Finally, we reformulate the fermions coupled to the gauge field on the graph as the covering graph derived from the gauge group.
Finally, we formulate gauge theory including fermions on the graph from the viewpoint of the covering graph derived from the gauge group in a unified way.

\newpage

\section{Introduction}
\label{sec:intro}

The construction of fermions in discrete space-time is important not only for understanding the behavior of electrons in crystals in condensed matter theory, but also for describing matter fields in gauge theories regularized on a lattice.
%There are different difficulties in defining fermions in discrete space than in a space-time continuum. 
%In particular, the fermion doubling problem appears when constructing chiral fermions on a lattice \cite{Nielsen:1980rz,Nielsen:1981xu}. This problem is overcome by a variety of ideas, which also bring a novel mathematical structure not found in the space-time continuum
%On the other hand, from a mathematical point of view, there are attempts to understand statistical models such as the Ising and dimer models in terms of fermions on the discrete spaces, which suggests a connection with integrable systems 
Defining fermions in discrete space-time presents unique challenges compared to the continuum space-time. 
Notably, the fermion doubling problem arises when constructing chiral fermions on a lattice \cite{Nielsen:1980rz,Nielsen:1981xu}. 
This issue is addressed through various innovative approaches, introducing novel mathematical structures absent in the space-time continuum \cite{Ginsparg:1981bj,Bodwin:1987ah,Neuberger:1997fp,Neuberger:1998wv}. 
Additionally, from a mathematical perspective, there are efforts to interpret statistical models, such as the Ising and dimer models, in terms of fermions in discrete spaces, suggesting a link to integrable systems \cite{Kenyon2002TheLA}.

The authors recently have constructed gauge theories on an arbitrary graph, i.e.~on a general discrete space-time, and studied their properties 
in %\cite{Ohta:2021jze,
\cite{Matsuura:2022dzl,Matsuura:2022ner,Matsuura:2023ova,Matsuura:2024gdu,Matsuura:2024rcv}.
%\cite{Matsuura:2022dzl}. 
This gauge theory can be regarded as a generalization of the Kazakov-Migdal model \cite{kazakov1993induced} to general graphs, and the partition function of the gauge theory is expressed in terms of the Ihara zeta function \cite{Ihara:original} 
and the Bartholdi zeta function \cite{2000math.....12161B}
by suitably choosing the coupling constants \cite{Matsuura:2022dzl,Matsuura:2022ner}. 
%By suitably choosing the coupling constants, the gauge theory on the graph is associated with the so-called Ihara zeta function \cite{Ihara:original} and generalizes the Kazakov-Migdal model.
%The Ihara zeta function appearing in the model can be further modified to the Bartholdi zeta function \cite{2000math.....12161B}, which includes extra parameter counting bumps of the cycles on the graph \cite{Matsuura:2022ner}.
These graph theoretical analogs of the Riemann zeta function are collectively referred to as the graph zeta function.
%The gauge theory on the graph has a nice property that the partition function is expressed in terms of the graph zeta function, and 
We can show attractable physical and mathematical properties like phase structure and dualities (functional equations) of the model by using the nature of the graph zeta function \cite{Matsuura:2023ova,Matsuura:2024gdu}. 
Correspondingly, a structure of the poles of the Bartholdi zeta function is studied in detail to understand the phase structure of the theory in more general parameter region \cite{Matsuura:2024rcv}.

In this paper, we propose a novel model of fermions on the graph. %whose partition function reproduces the graph zeta function, 
%and discuss its properties. 
%While the partition function of the bosonic model is expressed in terms of the graph zeta function, the partition function of the fermionic model is given by its inverse.
%The fermion on the graph described by graph zeta functions 
This model possesses a number of interesting properties. 
First, 
the fermions are defined both on the vertices and edges of the graph and the partition function of the model is descrived by the inverse of the graph zeta function.  %The graph is a set of vertices and edges, and the fermions are defined on the vertices and edges.
The Dirac operator of the fermions consists of deformed incidence matrices, 
which are regarded as first order difference operators on the graph, 
and mass terms. 
These fermions have no species doublers for the same reason that the staggered fermions do not.
In addition, the Dirac operator possesses the so-called $\gamma_5$-hermiticity, which allows us to construct the overlap fermion on the graph.
Moreover, by using the relationship between the graph zeta function of the covering graph and the Artin-Ihara $L$-function, 
we can construct the model on the so-called grid graphs including the square and honeycomb lattices. 
Applying this construction of the fermionic model to the two-dimensional grid, 
we can reproduce the phase transition point of the two-dimensional Ising model on the grid 
from the distribution of the poles of the Artin-Ihara $L$-function.
We also point out that, by introducing gauge fields on the graph, 
the partition function of the model becomes the inverse of the unitary matrix weighted graph zeta function
% which is the same as the (inverse of) the partition function of 
appeared in the generalized Kazakov-Migdal model. 
From the construction, the unitary matrix weighted graph zeta function should be regarded as an Artin-Ihara $L$-function of the covering graph constructed by the gauge group rather than the weighted graph zeta function.
%(graph zeta function of the derived covering graph of the gauge group). 

The paper is organized as follows. 
In the section \ref{sec:free fermion}, we introduce a deformation of the incidence matrix and the graph Laplacian on the graph. 
After discussing the properties of the free bosons and the free fermions on the graph, 
we introduce a fermionic model whose partition function is expressed in terms of the inverse of the graph zeta function.
In the section \ref{sec:properties}, we discuss the properties of the model and the meaning of the partition function. 
In this perspective, we introduce the concept of the fermionic cycles, 
which gives an interpretation of the coefficients of the inverse of the graph zeta function (a finite polynomial) as the number of the cycles with fermionic nature. 
In the section \ref{sec:grid graph}, we discuss the fermionic model on grid graphs by using the covering graph and the Artin-Ihara $L$-function. 
In the section \ref{sec:winding}, we discuss the relationship between the fermions on the graph and the two-dimensional Ising model and show that the poles of the graph zeta function determine the phase transition point. 
In the section \ref{sec:gauge}, we discuss the interacting fermion model with the gauge field on the graph and show that the partition function of the model is expressed in terms of the graph zeta function of the covering graph derived from the gauge group. 
In the section \ref{sec:conclusion}, we summarize our results and discuss future directions.

\section{Free Fermion on the Graph}
\label{sec:free fermion}

\subsection{Incidence matrix and Dirac operator}

A graph $\Gamma=(V,E)$ consists of vertices and edges,
where vertices are connected by edges.
We here consider a connected graph and denote a set of the vertices
and edges by $V$ and $E$, respectively.
The number of the vertices and edges are denoted by
$n_V=|V|$ and $n_E=|E|$. 
We only consider the directed graph in the following, 
where each edge has a direction and we can regard the edge as an arrow beginning from a vertex and ending to another vertex. 
%set the beginning of an edge from one vertex and the end of an edge at another vertex.

The incidence matrix for the directed graph is defined by
\be
{L^e}_v=
\begin{cases}
1 & \text{if $v=t(e)$}\\
-1 & \text{if $v=s(e)$}\\
0 & \text{others}
\end{cases},
\ee
where $s(e)$ and $t(e)$ represents
the vertex at the beginning (``source'')
and the vertex at the end (``target'')
of the edge $e$, respectively.
We can regard the incidence matrix as a first order difference operator,
which acts on a vector space $x^v$ on $V$ like
\be
{L^e}_v x^v = x^{t(e)}-x^{s(e)}.
\ee
For later convenience, we also introduce source and target matrices
as 
\be
{S^e}_v=
\begin{cases}
1 & \text{if $v=s(e)$}\\
0 & \text{others}
\end{cases},\quad
{T^e}_v=
\begin{cases}
1 & \text{if $v=t(e)$}\\
0 & \text{others}
\end{cases}.
\ee
Using them, the incidence matrix can be written by
\be
L = T-S\, .
\label{L by T and S}
\ee

The square of the incidence matrix,
\be
\Delta = L^T L\,,
\label{Laplace and Incidence}
\ee
is called the Laplacian matrix on the graph, 
%$\Delta$ 
since it acts on a vector $\boldsymbol{x}=(x^1,x^2,\cdots,x^{n_V})^T$ on $V$ as 
a second order difference operator
\be
\boldsymbol{x}^T L^T L \boldsymbol{x}
=\sum_{e\in E}(x^{t(e)}-x^{s(e)})^2\,.
\ee
%Thus, the graph Laplacian $\Delta$ is given by the incidence matrix as
The graph Laplacian is also represented by
\be
\Delta = D-A,
\label{eq:D-A}
\ee
where
$D$ is a diagonal matrix called the degree matrix whose diagonal elements are given
by the degree of each vertex, 
i.e.~the number of the edges connected to the vertex, 
and $A$ is the adjacency matrix defined by
\be
{A^v}_{v'}=\{
\text{the number of edges connecting 
adjacent (neighbor) vertices 
$v$ and $v'$}\}\, .
\ee
%Using the above expression of the incidence matrix
%by the source and target matrices
%in eq.~(\ref{L by T and S}),
%Since the incidence matrix is expressed by the source and target matrices as 
%eq.~(\ref{L by T and S}),

Using the incidence matrix (\ref{L by T and S}) expressed by the source and target matrices,
the graph Laplacian is written as 
\be
L^T L = (T^T-S^T)(T-S)= (T^TT+S^TS)-(T^TS+S^TT)\,. 
\ee
Then, comparing it to the expression \eqref{eq:D-A}, we see 
\be
\begin{split}
D&= T^TT+S^TS,\\
A&=T^TS+S^TT\,.
\end{split}
\ee

Using the relation (\ref{Laplace and Incidence}), we find
\be
\Ker \Delta = \Ker L,
\ee
since $\boldsymbol{x}^T \Delta \boldsymbol{x}=|L \boldsymbol{x}|^2$.
% We assume that the graph $\Gamma$ is connected in the following arguments.
Then, 
we can easily show that $\dim \Ker L = 1$ and thus $\dim \Ker \Delta =1$ ($\rank \Delta =n_V-1$). 
In fact, 
if $\boldsymbol{x}\in \Ker L$, 
$\boldsymbol{x}$ satisfies 
$x^{t(e)}=x^{s(e)}$ for $\forall e\in E$
over the connected part of the graph.
Since we assume the graph is connected, all elements of the vector $\boldsymbol{x}$ must have the same value. 
Therefore, $\boldsymbol{x} \in \Ker L$ is a ``constant mode'' $\boldsymbol{x}=c(1,1,\cdots,1)^T$ with a constant $c$ and $\dim \Ker L =1$.
%Apparently, this is a unique element of $\Ker L$ on the connected graph.
In particular, 
when $n_V\leq n_E$, 
we find $\dim \Ker L^T=n_E-n_V+1$.

%(We can discuss more on the fermion associated with $L$.)
%In the next subsection, 
Let us now consider ``field theories'' on the graph. 
As a first trial, we put bosonic degrees of freedom on vertices $v\in V$ 
which are expressed in terms of an $n_V$-dimensional vector on $V$;
$\bphi=(\phi^1,\phi^2,\cdots,\phi^{n_V})^T$.
If we regard it as a massless scalar field, 
a natural action on the graph is defined through the graph Laplacian 
%The action for a massless scalar field is defied by the graph Laplacian
as
\be
S_B = \bphi^T\Delta \bphi\, .
\ee
The partition function for this model is given by integration over the vector $\bphi$, 
\be
Z_B = \int \prod_{v\in V} d\phi^v \, e^{-\beta S_B},
\label{bosonic path integral}
\ee
where $\beta$ is an overall coupling constant.
Since the partition function (\ref{bosonic path integral}) is essentially
Gaussian, we can estimate the partition function as 
\be
Z_B = \left(\frac{2\pi}{\beta}\right)^{n_V-1}\int d\phi_0 \, \frac{1}{\sqrt{\det' \Delta}},
\ee
where $\phi_0$ is one zero mode in $\Ker \Delta$ and $\det' \Delta$ stands
for the determinant of the Laplacian without the zero mode
(the product of the non-zero eigenvalues of $\Delta$).
Due to the existence of the bosonic zero mode,
the partition function $Z_B$ diverges in general.
%So we need to insert a suitable observables to regularize
%the integral over the zero mode.
So we need to insert a suitable observable to regularize the zero mode in order to make the model well-defined.

%What would be the case with fermions?
We next try to put massless fermions on the graph. 
%As we discussed above, since the graph Laplacian is associated with the
%second order difference operator,
%the kinetic term of the fermionic fields on the graph should be written by the incidence matrix.
From the nature of the fermions, 
the kinetic term of the fermion should be written by the incidence matrix 
since the incidence matrix is associated with the first order difference operator as discussed above. 
Since the incidence matrix is $n_E \times n_V$ matrix,
we need to introduce not only the fermions on the vertices $V$
but also the fermions on the edges $E$.
Then, 
if we denote the fermions (Grassmann variables) on $V$ and $E$
as $\boldsymbol{\xi}=(\xi^1,\xi^2,\cdots,\xi^{n_V})^T$
and $\boldsymbol{\psi}=(\psi^1,\psi^2,\cdots,\psi^{n_E})^T$, 
respectively, 
the fermionic action is invoked 
\be
S_F = \boldsymbol{\psi}^T L\, \boldsymbol{\xi}-\boldsymbol{\xi}^T L^T \boldsymbol{\psi}.
\ee
The partition function for this model is given by
\be
Z_F = \int \prod_{v\in V}d\xi^v \prod_{e\in E}d\psi^e \,
e^{-\beta S_F}\, ,
\ee
which is again 
%So this partition function is 
a Gaussian integral for the Grassmann variables. 
Then, 
we can evaluate the partition function as 
\be
Z_F =\int d\xi_0 \prod_{i=1}^{n_E-n_V+1} d\psi_0^i \,
\Pf' \begin{pmatrix}
  0 & -L^T \\ L & 0 
\end{pmatrix}\, ,
\ee
where $\xi_0$ and $\psi_0^i$ ($i=1,\cdots,n_E-n_V+1$)
are zero modes ($\Ker L$ and $\Ker L^T$)
%and $\Pf'$ is a Pfaffian restricted to the non-zero modes.
%For the fermionic case, 
%In this case, 
%the partition function vanishes
%due to the existence of the zero modes. 
%Thus, in order to make the theory well-defined, 
%So if we would like to obtain a non-vanishing physical observable,
and $\Pf'$ is a Pfaffian restricted to the non-zero modes.
%For the fermionic case, 
In this case, 
the partition function vanishes
due to the existence of the zero mode. 
Thus, in order to make the theory well-defined, 
%So if we would like to obtain a non-vanishing physical observable,
we need to insert a suitable fermion zero mode operator like
${\cal O}_0=\psi_0\prod_{i=1}^{n_E-n_V+1} \xi_0^i$.

%The partition function defined above is one of a trial for graph theoretical fermion model, but it seems to be too naive.
%These two simple examples show that naive field theories on the graph suffer from the zero modes in general. 
%In the following, we overcome this problem by considering a more sophisticated fermionic model associated with the graph zeta function.
%We will see that the model has quite preferable
%properties not only for mathematics but also for physics.
These two simple examples show that naive field theories on the graph suffer from the existence of the zero modes in general. 
In the following, we overcome this problem by considering a more sophisticated fermionic model associated with the graph zeta function.
We will see that the model has quite preferable
properties not only for mathematics but also for physics.

\subsection{Deformed graph Laplacian and graph zeta function}
\label{subsec:deformed Laplacian}

Now let us consider a fermionic model whose partition function is given by 
Bartholdi's graph zeta function, which is an 
extension of the Ihara zeta function.
To define the graph zeta function, 
%we need to introduce the concept of the primitive cycle of the graph.
we need to explain the concept of cycles of the graph.

First, we introduce a set of the directed edges $E$ and their inverses $\bar{E}$.
The inverse edge $\bar{e}=\langle w,v\rangle$ has reversed direction of the edge $e=\langle v,w\rangle$.
Since the directed edges have always paired inverse edges, we find $|E|=|\bar{E}|$. Then 
we can combine them to a set of the undirected edges $E_D=E\cup \bar{E}$ of $|E_D|=2n_E$.
We denote the elements of $E_D$ by
\be
E_D = \{\bse_1,\bse_2,\cdots,\bse_{2n_E}\}\equiv  \{e_1,\cdots,e_{n_E},\bar{e}_1,\cdots,\bar{e}_{n_E}\}.
\ee

Secondly, a path $P$ on the graph is given by a sequence of the edges in $E_D$ such that $P=\bse_1\bse_2\cdots \bse_{k}$
satisfying $t(\bse_i)=s(\bse_{i+1})$ for $i=1,2,\cdots,k-1$, where $k$ is called the length of the path.
If a path $P=\bse_1\bse_2\cdots \bse_{k}$ satisfies $t(\bse_k)=s(\bse_1)$, then the path is called a cycle $C$
of length $k$, which is denoted by $\ell(C)$.
A cycle $C$ is called {\em primitive} if it is not 
expressed as a concatenation of the two or more same cycles,
that is
$C\neq (C')^r$ ($r \geq 2$) for any cycle $C'$.
A part of a cycle $C=\bse_1\bse_2\cdots \bse_{k}$ is called a bump
if $\bse_i=\bar{\bse}_{i+1}$ ($i=1,\cdots,k-1$) for some $i$ or $\bse_k=\bar{\bse}_1$.
The number of the bumps in the cycle $C$ is called the cyclic bump count and denoted by $b(C)$.
Two cycles $C=\bse_1\bse_2\cdots \bse_{k}$ and $C'=\bse'_1\bse'_2\cdots \bse'_{k}$
with the same length are called equivalent if $\bse'_i=\bse_{i+j}$ for some $j$.
So we can define the equivalence class $[C]$ of the cycle $C$.

Under these preparations, we define the Bartholdi zeta function of the graph by the Euler product
\be
\zeta_\Gamma(q,u)
=\prod_{[C]:\text{primitive}}\frac{1}{1-u^{b(C)}q^{\ell(C)}},
\label{Bartholdi zeta function}
\ee
where $[C]$ runs over all equivalence classes of primitive cycles on $\Gamma$.
This is a generating function of the number of the cycles as a power series of $q$ and $u$.
Taking $u=0$,
the factors including non-zero bump counts is dropped in the product and
the Bartholdi zeta function reduces to the Ihara zeta function
\be
\zeta_\Gamma(q,u=0)
=\prod_{[C]:\substack{\text{primitive}\\ \text{reduced\ \,}}}\frac{1}{1-q^{\ell(C)}}\,,
\ee
where $[C]$ now runs over all equivalence classes of primitive cycles without bumps
(primitive and reduced cycles) on $\Gamma$.

The Bartholdi zeta function can be written in terms of a determinant of the deformed
graph Laplacian as
\be
\zeta_\Gamma(q,u) \equiv
\frac{1}
{(1-q^2(1-u))^{n_E-n_V}
\det \Delta_{q,u}
}\,.
\label{eq:vetex representation of Bartholdi}
\ee
Here, $\Delta_{q,u}$ is a two parameter deformation of the graph Laplacian
defined by
\be
\Delta_{q,u} \equiv I_{n_V}-q A +q^2(1-u)\left(D-(1-u)I_{n_V}\right),
\label{deformed Laplacian}
\ee
where $I_{n_V}$ is an $n_V \times n_V$ identity matrix.
By setting $q=1$ and $u=0$, the deformed graph Laplacian reduces to
the original graph Laplacian $\Delta=D-A$.

%Interestingly,
The Bartholdi zeta function has another expression called the Hashimoto expression
\be
\zeta_G(q,u)=\frac{1}{\det\left(I_{2n_E}-qB_u\right)},
\label{Hashimoto expression}
\ee
where $B_u\equiv W+uJ$, and
$W$ and $J$ are $2n_E\times 2n_E$ matrices defined by
\be
W_{\bse \bse'} = \begin{cases}
1 &\text{if $t(\bse')=s(\bse)$ and $\bse'\neq \bar{\bse}$}\\
0
\end{cases},
\quad
J_{\bse \bse'} = \begin{cases}
  1 &\text{if $\bse' = \bar{\bse}$}\\
  0
  \end{cases}\,.
\ee
$W$ is called the edge adjacency matrix
which can be regarded as the adjacency matrix of the oriented line graph derived from $\Gamma$.
Using $S$ and $T$, we can express $W$ as a blockwise matrix
\be
W = \begin{pmatrix}
  TS^T & TT^T-I_{n_E}\\
  SS^T-I_{n_E} & ST^T
\end{pmatrix}\, .
%\quad
%J=\begin{pmatrix}
%0 & I_{n_E}\\
%I_{n_E} & 0
%\end{pmatrix}\, .
\ee
$J$ is a matrix with an off-diagonal block of size $n_E$ identity matrices,
whose non-vanishing element makes $e_i$ and $\bar{e}_i$ adjacent and creates a bump.
This is the reason why the parameter $u$, which appears in front of $J$ of $B_u$,
counts the number of the bumps.

Note that, 
since the Bartholdi zeta function is written in terms of the
deformed graph Laplacian,
we can regard the Bartholdi zeta function as a partition function
%of a model of 
a bosonic model with the scalar field $\bphi$ on $V$
%, which is
%now taken as an $n_V$-dimensional complex vector to obtain the determinant.
%With an action
with the action, 
\be
S_B(q,u) = \boldsymbol{\phi}^\dag\Delta_{q,u}\boldsymbol{\phi}\,. 
\ee
In fact, 
the partition function (Gaussian integral) of the bosonic model reduces to
%\be
%\begin{split}
\begin{align}
Z_B(q,u) &= \int \prod_{v\in V}d\phi^v d\phib^v\,
e^{-\beta S_B(q,u)} \nn \\
&=\left(\frac{2\pi}{\beta}\right)^{n_V}\frac{1}{\det \Delta_{q,u}} \nn \\
&= \left(\frac{2\pi}{\beta}\right)^{n_V} (1-q^2(1-u))^{n_E-n_V} \zeta_\Gamma(q,u)\,,
\label{zeta by boson}
%\end{split}
%\ee
\end{align}
where we have used the relation (\ref{eq:vetex representation of Bartholdi}).
%which gives the Bartholdi zeta function itself by choosing the coupling constant
%$\beta$ suitably.

%In this bosonic model, there is no subtle problem caused by zero mode, 
This bosonic model does not suffer from the zero mode problem,
since the deformed graph Laplacian has essentially a mass term and zero modes are uplifted. 
Note that the generalized Kazakov-Migdal model \cite{Matsuura:2022dzl,Matsuura:2022ner,Matsuura:2023ova,Matsuura:2024gdu,Matsuura:2024rcv} can be regarded as an extention of this bosonic model to the gauge theory on the graph.

%The bosonic model associated with the graph zeta function can be extended to gauge theory including non-Abelian gauge field on the edges \cite{Matsuura:2022dzl,Matsuura:2022ner,Matsuura:2023ova,Matsuura:2024gdu,Matsuura:2024rcv},
%which can be understood as a generalization of the Kazakov-Migdal model \cite{kazakov1993induced} on the graph.

\subsection{Fermion associated with the zeta function on the graph}
\label{sec:fermion theory}

Let us now consider a fermionic model associated with the Bartholdi zeta function.
To obtain the deformed graph Laplacian in the fermionic model as a functional determinant,
we need an appropriate Dirac operator. 
Since the Dirac operator should be written in terms of the first order difference operator,
it is useful to define deformed forward and backward difference operators
(incidence matrix) as
\be
L_{q,u} \equiv T-tS, \quad \Lt_{q,u} \equiv S-tT\, ,
\ee
where we have defined $t= q(1-u)$.
They reduce to $L_{q,u}=-\Lt_{q,u}=L$ when $t=1$.
%It reduces to simply that $L_{q,u}=-\Lt_{q,u}=L$ 
%for $u=0$ and $q=1$, namely $t=1$.
Using these deformed incidence matrices, the deformed graph Laplacian $\Delta_{q,u}$ can be expressed as
\be
  \Delta_{q,u}
=\left(1-t^2\right)I_{n_V}-q S^T L_{q,u}-q T^T \Lt_{q,u} \ .
\ee
%where  is the two parameter deformation
%of the graph Laplacian (\ref{deformed Laplacian}).
% where
% \be
% \Delta_q \equiv I_{n_V}-qA+q^2 (D-I_{n_V}).
% \ee
% Thus
% \be
% \Delta_q +qS^TL_q+qT^T\Lt_q = (1-q^2)I_{n_V}
% \ee

\begin{figure}[t]
  \begin{center}
    \includegraphics[scale=0.9]{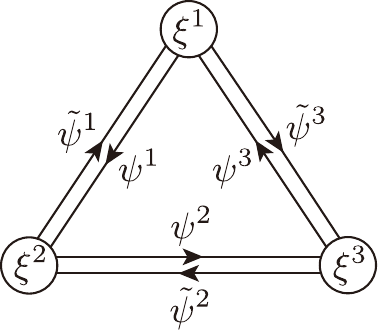}
  \end{center}
  \caption{The graph of the cycle graph $C_3$, which has three vertices and three edges. There are three fermions $\xi^v$ on each vertex $v$ and three pairs of fermions $(\psi^e,\tilde{\psi}^e)$ on each edge $e$.}
  \label{cycle graph C3}
\end{figure}

\begin{figure}[t]
  \begin{center}
    \includegraphics[scale=0.8]{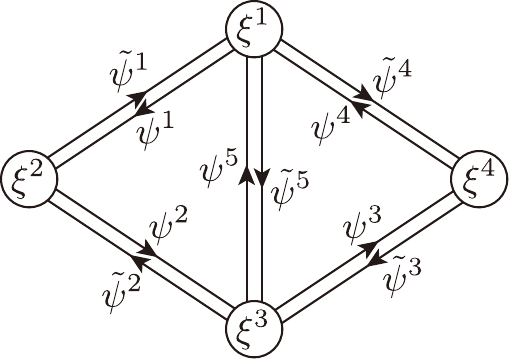}
  \end{center}
  \caption{The graph of the double triangle graph DT, which has four vertices and five edges. There are four fermions $\xi^v$ on each vertex $v$ and five pairs of fermions $(\psi^e,\tilde{\psi}^e)$ on each edge $e$.}
  \label{double triangle graph}
\end{figure}

%Using this kind of the first order
%difference operators $L_{q,u}$ and $\Lt_{q,u}$,
%we first try to construct the Dirac operator
%like
Combining the deformed incidence matrices, we define a Dirac operator as 
\be
\slashed{D} = \alpha\begin{pmatrix}
0 & \Lt_{q,u}^T & L_{q,u}^T\\
L_{q,u} & 0 & 0 \\
\Lt_{q,u} & 0 & 0
\end{pmatrix}\ ,
\ee
where $\alpha=\sqrt{\frac{q}{1-t^2}}$ is a normalization constant introduced for later convenience.
%The Dirac operator is acting on three component fermion
Corresponding to the structure of this operator, we introduce fermions
\be
\Psi=(\xi, \psi,\psit)^T,\qquad
\bar{\Psi} = (\bar{\xi},\bar{\psi},\bar{\psit}),
\label{eq:fermions on the graph}
\ee
where
$\xi^v$ and $(\psi^e,\psit^e)$ are Grassmann variables defined on $V$ and $E$, respectively, 
and $\bar\xi^v$ and $(\bar\psi^e,\bar\psit^e)$ are their complex conjugate.
Examples of the assignment of the fermions on two kinds of the graph
(cycle graph and double triangle graph) are shown in Fig.\ref{cycle graph C3} and Fig.\ref{double triangle graph}. We will use this assignment of the fermions on the graph through out the paper.

Using them, one may consider a model of massless fermions with the action
%If we consider these fermions massless, the action will be 
%An action for the massless fermion is expressed as
\be
\begin{split}
S_F(q,u) &= \bar{\Psi} \slashed{D} \Psi\\
&=\alpha\left\{
\bar{\psi}L_{q,u}\xi+\bar{\psit}\Lt_{q,u}\xi
+\bar{\xi}\Lt_{q,u}^T\psi+\bar{\xi}L_{q,u}^T \psit\right\}
\ .
\end{split}
\label{eq:fermion action}
\ee
%After integrating out the fermions in the path integral
%of the partition function,
%we obtain the determinant of the above Dirac operator,
%but 
However, this is not suitable for our purpose since the determinant of the operator $\slashed{D}$ obtained by integrating out the fermions is nothing to do with the deformed graph Laplacian. 
%it does not give $\det \Delta_{q,u}$.
%\be
%\slashed{D}^2 = \alpha^2\begin{pmatrix}
%  \Lt_{q,u}^TL_{q,u}+L_{q,u}^T \Lt_{q,u} & 0 & 0\\
%  0 & L_{q,u}\Lt^T_{q,u} & L_{q,u}L^T_{q,u} \\
%  0 & \Lt_{q,u}\Lt^T_{q,u} & \Lt_{q,u}L^T_{q,u}
%  \end{pmatrix}, 
%\ee
%and then
%the determinant of $\slashed{D}$ is not expressed in terms of the deformed graph Laplacian.
In order to reproduce it, 
we also have to introduce a mass operator
%In addition, we introduce the fermion mass as follows:
\be
{\cal M} =
\begin{pmatrix}
  I_{n_V} & 0 & 0 \\
0 & I_{n_E} & -tI_{n_E}\\
0 & -tI_{n_E} & I_{n_E}
\end{pmatrix}\,, 
\ee
%Then, the action for a massive fermion is modified to
and consider massive fermions with the action
\be
S_F(q,u) = \bar{\Psi}\left(\slashed{D}+{\cal M}\right)\Psi\, ,
\label{fermion action}
\ee
where
\be
\slashed{D}+{\cal M} =
\begin{pmatrix}
I_{n_V} & \alpha\Lt_{q,u}^T & \alpha L_{q,u}^T\\
\alpha L_{q,u} & I_{n_E} & -tI_{n_E}\\
\alpha \Lt_{q,u} & -tI_{n_E} & I_{n_E}
\end{pmatrix}\, .
\ee
%If we consider the partition function of this massive fermion, we obtain
The partition function of the model is given by 
\be
Z_F(q,u) = {\cal N}\int \D \Psi \D \bar{\Psi} \, e^{-\beta S_F(q,u)}
={\cal N}\beta^{n_V+2n_E}\det \left(\slashed{D}+{\cal M}\right),
\ee
where $\beta$ is an overall coupling constant and ${\cal N}$ is a normalization constant
of the path integral measure.
To evaluate the determinant %of this partition function 
in this expression explicitly, 
%we first consider the following matrix decomposition
it is useful to decompose the matrix $\slashed{D}+{\cal M}$ as 
\be
\slashed{D}+{\cal M} =
\begin{pmatrix}
  I_{n_V} & \alpha S^T &  \alpha T^T\\
  0 & I_{n_E} & 0\\
  0 & 0 & I_{n_E}
\end{pmatrix}
\begin{pmatrix}
\frac{\Delta_{q,u}}{1-t^2} & 0 & 0\\
\alpha L_{q,u} & I_{n_E} & -tI_{n_E} \\
\alpha \Lt_{q,u} & -tI_{n_E} & I_{n_E}
\end{pmatrix}
\, .
\label{eq:matrix decomposition}
\ee
Then, the determinant of the Dirac operator can be evaluated as 
\be
\det \left(\slashed{D}+{\cal M}\right)
=(1-t^2)^{n_E-n_V}\det \Delta_{q,u}=\zeta_\Gamma(q,u)^{-1}
\, ,
\ee
as announced, 
and 
%Thus, we find that 
the partition function of the massive fermion can be written
in terms of the inverse of the Bartholdi zeta function as
\be
Z_F(q,u) = {\cal N}\beta^{n_V+2n_E}\zeta_\Gamma(q,u)^{-1}\,.
\ee
In particular, 
it becomes the inverse of the Bartholdi zeta function itself %as the partition function of the massive fermion
by tuning the coupling constant $\beta$ and normalization constant ${\cal N}$ suitably. 
%\footnote{We can choose $\beta=q$ and ${\cal N}=(q-q^3(1-u)^2)^{-n_V}$ for example.},
%\be
%Z_G(q,u) = \zeta_G(q,u)^{-1}.
%\ee
%With the understanding of this normalization, we will consider the partition function of this free fermion model.

More interestingly, $\slashed{D}+{\cal M}$ has another decomposition
\be
\slashed{D}+{\cal M}
 =
\left(
%\begin{array}{c|c}
%I_{n_V} & \begin{array}{cc}0\quad\ & \quad 0\end{array}\\
%\hline
%\alpha L_{q,u} & \\
 %& \dfrac{\left(I_{2n_E}-t J\right)
 %\left(I_{2n_E}-qB_u\right)}{1-t^2}
 %\\
%\alpha \Lt_{q,u} &
%\end{array}
%\right)
\begin{array}{c|c}
I_{n_V} & 0\hspace{2cm} 0  \\
\hline
\begin{array}{c} \alpha L_{q,u} \\ \alpha \Lt_{q,u} \end{array}
& 
\dfrac{\left(I_{2n_E}-t J\right) \left(I_{2n_E}-qB_u\right)}{1-t^2}
\end{array}
\right)
%\times
\begin{pmatrix}
  I_{n_V} & \alpha \Lt^T_{q,u} &  \alpha L^T_{q,u}\\
  0 & I_{n_E} & -t I_{n_E} \\
  0 & -t I_{n_E} & I_{n_E}
  \end{pmatrix}\, ,
\ee
%\begin{pmatrix}
  %(1-t^2q^2)\bs{1}_{Kn_V} &  q(1-t^2q^2)S^T \hspace{2cm} q(1-t^2q^2)T^T \\
  %\begin{array}{c} 0 \\ 0 \end{array} &
  %\left(\bs{1}_{2Kn_E}-q(W_X+(1-t)J_X)\right)
  %\left(\bs{1}_{2Kn_E}-tq J_X\right)
%\end{pmatrix}\,.
%So we find another expression of the determinant of the Dirac operator,
which yields the determinant of the Dirac operator 
associated with the Hashimoto expression \cite{Hashimoto1990ONZA}, 
\be
\det \left(\slashed{D}+{\cal M}\right) =
\det\left(I_{2n_E}-qB_u\right)\, ,
\ee
since $\det(I_{n_E}-t J)=(1-t^2)^{n_E}$.
This equivalence of the two representations of the fermion determinant also shows that the equivalence of the Ihara and Hashimoto expressions
\cite{bass1992ihara}
\be
\begin{split}
\zeta_\Gamma(q,u)^{-1}&=
(1-t^2)^{n_E-n_V}\det \Delta_{q,u}\\
&=\det\left(I_{2n_E}-qB_u\right)\, .
\end{split}
\ee

\section{Properties of the Fermionic Partition Function}
\label{sec:properties}

In this section, we discuss the meaning of the fermionic model constructed in the previous section whose partition function 
%fermion partition function which 
is expressed in terms of the inverse of the Bartholdi zeta function.

\subsection{Infinite product expansion}

From the definition of the Bartholdi zeta function (\ref{Bartholdi zeta function}),
its inverse is also expressed by a product, 
\be
\zeta_\Gamma(q,u)^{-1}
=\prod_{[C]:\text{primitive}}(1-u^{b(C)}q^{\ell(C)})\, .
\label{inverse of Bartholdi zeta function}
\ee
Since there are infinitely many primitive cycles on the graph in general,
this is an infinite product.
Taking the logarithm of (\ref{inverse of Bartholdi zeta function}), we find
\begin{align}
%\begin{split}
\log \zeta_\Gamma(q,u)^{-1} 
&= \sum_{[C]:\text{primitive}}\log(1-u^{b(C)}q^{\ell(C)}) \nn \\
&= -\sum_{[C]:\text{primitive}}\sum_{k=1}^\infty \frac{\left(u^{b(C)}q^{\ell(C)}\right)^k}{k} \nn \\
&= -\sum_{C:\text{primitive}} \sum_{k=1}^\infty \frac{u^{b(C^k)}q^{\ell(C^k)}}{\ell(C^k)}\nn \\
&= -\sum_C \frac{u^{b(C)}q^{\ell(C)}}{\ell(C)}\nn \\
&= -\sum_{\ell = 1}^\infty \frac{N_\ell(u)}{\ell}q^\ell
\, ,
\label{From Bartholdi}
%\end{split}
\end{align}
where we have used, in the third equality, the fact that there are $\ell(C)$ elements in the equivalence class
by changing the sum from $[C]$ to $C$,
and $b(C^k)=kb(C)$ and $\ell(C^k)=k\ell(C)$.
%At $u=0$, $N_\ell(0)$ stands for the number of the reduced cycles with length $\ell$.
%Now $N_\ell(u)$ is a function of $u$ and has the series expansion in $u$ as
The coefficient $N_\ell(u)$ in \eqref{From Bartholdi} can be expanded in terms of $u$ as
\be
N_\ell(u)
=\sum_{b\geq 0} N_{\ell,b}u^b\, ,
\ee
where $N_{\ell,b}$ is the number of the cycles of length $\ell$ with $b$ bumps including the cardinality of the equivalence class.
Note that $N_\ell(u)$ becomes the number of the reduced (but not need to be primitive) cycles with length $\ell$ at $u=0$.

On the other hand,
the product \eqref{inverse of Bartholdi zeta function} can be arranged as 
\be
%\begin{split}
\zeta_\Gamma(q,u)^{-1}
%&=\prod_{[C]:\text{primitive}}(1-u^{b(C)}q^{\ell(C)})\nn \\
=\prod_{\ell=1}^\infty\prod_{b=0}^\infty(1-u^bq^\ell)^{\pi_{\ell,b}}\,,
%\end{split}
\label{infinite product}
\ee
where $\pi_{\ell,b}$ stands for the multiplicity of the cycles of length $\ell$ with $b$ bumps in the equivalence class of the cycles.
Taking the logarithm of the expression (\ref{infinite product}),
we obtain
\begin{align}
%\begin{split}
\log \zeta_\Gamma(q,u)^{-1}
&=\sum_{\ell=1}^\infty\sum_{b=0}^\infty \pi_{\ell,b}\log(1-u^bq^\ell)\nn \\
&=-\sum_{\ell=1}^\infty\sum_{b=0}^\infty \sum_{k=1}^\infty \frac{\pi_{\ell,b}u^{kb}q^{k\ell}}{k}\nn \\
&=-\sum_{n=1}^\infty\sum_{d|n}\sum_{b=0}^\infty \frac{\pi_{d,b}u^{nb/d}q^n}{n/d}\nn \\
&=-\sum_{n=1}^\infty  \frac{\sum_{d|n}d \, \pi_d(u^n)}{n}q^{n}\,,
%\end{split}
\label{From infinite product}
\end{align}
where
\be
\pi_d(u) = \sum_{b\geq 0} \pi_{d,b}u^{b/d}\,.
\ee
Comparing (\ref{From infinite product}) with (\ref{From Bartholdi}),
we find
\be
N_n(u^{1/n}) = \sum_{d|n}d \, \pi_d(u)\, .
\ee
Using the M\"obius inversion formula, we can express $\pi_d(u)$ in terms of $N_n(u)$ as
\be
\pi_\ell(u)=\frac{1}{\ell}\sum_{d|\ell}\mu\left(\frac{\ell}{d}\right)N_d(u^{1/d})\, ,
\ee
where $\mu(n)$ is the M\"obius function defined by
\be
\mu(n) = \begin{cases}
  1 & \text{if $n=1$}\\
  (-1)^p & \text{if $n$ is a product of $p$ distinct primes}\\
  0 & \text{if $n$ has a squared prime factor}\\
\end{cases}.
\ee

%It is very surprising, but according to Ihara's theorem \cite{Ihara:original}, 
%Surprisingly, 
The inverse of the graph zeta function reduces to a polynomial of finite degree of order $2n_E$
as a consequence of Ihara's theorem \cite{Ihara:original}
and the equivalent Hashimoto expression \cite{Hashimoto1990ONZA}, 
despite having an infinite product representation like (\ref{infinite product}).
This means that the partition function of our model 
%on the graph is expressed as both 
has two equivalent but seemingly different expressions in terms of an infinite product expansion and a finite series expansion up to the order $2n_E$.
Thus, we can extract $N_\ell(u)$ by evaluating the series expansion of the logarithm of the polynomial and determine each $\pi_{\ell,b}$ explicitly.

Let us check the above properties of the fermionic partition function
for concrete examples of the cycle graph $C_3$ and the double triangle graph (DT).
The cycle graph $C_3$ depicted in Fig.~\ref{cycle graph C3} contains three vertices and three edges.
For the cycle graph $C_3$, the inverse of the Bartholdi zeta function is given by
\be
\zeta_{C_3}(q,u)^{-1}=1-3 u^2 q^2 -2 q^3-\left(3 u^2-3 u^4\right)q^4 +\left(1-3u^2+3 u^4-u^6\right)q^6 \, .
\label{C3 inverse Bartholdi}
\ee
From the series expansion of the logarithm of the zeta function,
we see
\be
\begin{split}
&N_2(u)=6u^2,\quad
N_3(u)=6,\quad
N_4(u)=12 u^2 + 6 u^4,\quad
N_5(u)=30  u^2,\\
&N_6(u)=6 + 18 u^2 + 36 u^4 + 6 u^6,\quad
\cdots\,.
\end{split}
\ee
Using the M\"obius inversion formula, we obtain
\be
\begin{split}
&\pi_2(u)=3u,\quad
\pi_3(u)=2,\quad
\pi_4(u)=3u^{1/2},\quad
\pi_5(u)=6  u^{2/5},\\
&\pi_6(u)=3 u^{1/3} + 6 u^{2/3},\quad \cdots\,.
\end{split}
\ee
Then, picking up the coefficients of terms in $\pi_d(u)$, we find
\be
\pi_{2,2}=3,\quad
\pi_{3,0}=2,\quad
\pi_{4,2}=3,\quad
\pi_{5,2}=6,\quad
\pi_{6,2}=3,\quad \pi_{6,4}=6,
\quad \cdots\,.
\ee
Therefore, 
the infinite product expression 
of the inverse of the Bartholdi zeta function of the cycle graph $C_3$ becomes 
\be
\begin{split}
\zeta_{C_3}(q,u)^{-1}
&=\prod_{\ell=1}^\infty\prod_{b=0}^\infty(1-u^bq^\ell)^{\pi_{\ell,b}}\\
&=(1-u^2q^2)^3(1-q^3)^2(1-u^2q^4)^3\\
&\qquad\qquad\times
(1-u^2q^5)^6(1-u^2q^6)^3(1-u^4q^6)^6\cdots\, ,
\end{split}
\ee
which interestingly reduces to a polynomial of finite degree (\ref{C3 inverse Bartholdi}), 
that is, the terms with higher powers than $q^6$ are canceled out. 

The second example is the double triangle graph (DT), which
has four vertices and five edges as depicted in Fig.~\ref{double triangle graph}.
Since it is already cumbersome to write out all the terms including $u$ even for this DT case, we consider only the inverse of the Ihara zeta function by setting $u=0$,
\be
\zeta_{\rm DT}(q)^{-1}=1-4q^3-2q^4+4q^6+4q^7+q^8-4q^{10}\, .
\label{Inverse of Ihara for DT}
\ee
Using the same algorithm of the M\"obius inversion formula to find the powers of the infinite product, the infinite produce expression of the inverse of the Ihara zeta function of DT becomes 
\be
\begin{split}
\zeta_{\rm DT}(q)^{-1}
&=(1-q^3)^4(1-q^4)^2(1-q^6)^2(1-q^7)^4(1-q^9)^4\\
&\qquad\times(1-q^{10})^{12}
(1-q^{11})^4(1-q^{12})^6(1 - q^{13})^{32}(1 - q^{14})^{18}\cdots\,.
\end{split}
\ee
Again, the higher terms than $q^{10}$ in this expansion are canceled out. 

\subsection{Series expansion and fermionic cycles}
\label{sec:fermionic cycles}

%Using the Hashimoto expression of the Bartholdi zeta function \eqref{Hashimoto expression},
%we can immediately see that its inverse is 
As mentioned above, 
the inverse of the Bartholdi zeta function has a finite series of order $2n_E$ in $q$.
This means that the fermionic fields generate only a finite number of the cycles on the graph up to the length $2n_E$ due to the exclusion principle of the fermions, 
since the power of $q$ counts the number of cycles.

To see it more explicitly, we rewrite the inverse of the Bartholdi zeta function by fermion integral as 
\begin{align}
  \zeta_\Gamma(q,u)^{-1} &= \det(I_{2n_E}-qB_u) \nn \\
  &= \int \prod_{\bse\in E_D}d\eta_\bse d\bar\eta_\bse \, e^{\bar{\boldsymbol{\eta}} (1-qB_u)\boldsymbol{\eta}} \nn \\
  &= \int \prod_{\bse\in E_D}d\eta_\bse d\bar\eta_\bse\, \frac{1}{(2n_E)!}\left(\bar{\boldsymbol{\eta}} \left(I_{2n_E}-qB_u\right)\boldsymbol{\eta}\right)^{2n_E} \nn \\
  &= 
  \sum_{\ell=0}^{2n_E}\frac{(-q)^\ell}{\ell!(2n_E-\ell)!}
  \int \prod_{\bse\in E_D}d\eta_\bse d\bar\eta_\bse\, 
  \left(\bar{\boldsymbol{\eta}}\boldsymbol{\eta}\right)^{2n_E-\ell} \left(\bar{\boldsymbol{\eta}} B_u \boldsymbol{\eta}\right)^{\ell} 
  \label{eq:zeta_fermion tmp1}
\end{align}
where $\boldsymbol{\eta}=(\eta_{e_1}, \cdots,\eta_{e_{n_E}},\eta_{\bar{e}_1},\cdots, \eta_{\bar{e}_{n_E}})^T$ and $\bar{\boldsymbol{\eta}}=(\bar{\eta}_{e_1}, \cdots,\bar{\eta}_{e_{n_E}},\bar{\eta}_{\bar{e}_1},\cdots, \bar{\eta}_{\bar{e}_{n_E}})$ are independent $2n_E$-dimensional Grassmann valued vectors,
and we have used the nature of the Grassmann integral that the integrand must contain $2n_E$ Grassmann variables in the third line. 
Note that we have to normalize the measure of the Grassmann integral as 
\begin{align}
  \int \prod_{\bse\in E_D}d\eta_\bse d\bar\eta_\bse\, \prod_{\bse\in E_D} \bar\eta_\bse\eta_\bse = 1\,,
  \label{eq:fermion measure}
\end{align}
in order to hold \eqref{eq:zeta_fermion tmp1}. 
Since $\boldsymbol{\bar\eta\eta}$ consists of pairs of the Grassmann variables $\bar\eta_\bse \eta_\bse$ on the same edge $\bse$,
only such terms in the expansion of $\left(\bar{\boldsymbol{\eta}} B_u\boldsymbol{\eta}\right)^{\ell}$ 
that contain $\eta_{\bse_{i_1}}\cdots\eta_{\bse_{i_\ell}}$ and $\bar\eta_{\bse_{i_1}}\cdots\bar\eta_{\bse_{i_\ell}}$ of a common set of edges $\{\bse_{i_1},\cdots,\bse_{i_\ell}\}$ contribute to the integral.
From the definition of the matrix $B_u$, 
such edges must form a set of cycles on the graph. 
Furthermore, 
such a term that contributes to the integral cannot include the same $\eta_\bse$ and $\bar\eta_\bse$ twice or more. 
Thus, the cycles are all primitive and do not share the same edge with each other% 
\footnote{
  Note that we distinguish the inverse edge from an edge $e\in E$ as an different edge in this case since $\eta_{e}$ and $\eta_{\bar{e}}$ are independent Grassmann variables.
}.
Since primitive cycles made of the same edges form an equivalence class of cycles by identifying the cyclic rotation of the edges,
there is a one-to-one correspondence between a term in the expansion of $\left(\bar{\boldsymbol{\eta}} B_u\boldsymbol{\eta}\right)^{\ell}$ which contributes to the integral and a set of the equivalence classes of the primitive cycles of total length $\ell$. 
We call such a set of the equivalence classes as {\it a fermionic cycle} and denote it as $[\Psi]$.

Let us assume that a fermionic cycle $[\Psi]$ is made of $F$ equivalence classes of primitive cycles on the graph $\{[C_1],\cdots, [C_F]\}$ of length $\ell_i$ ($i=1,\cdots,F$) which satisfy $\ell_1+\cdots+\ell_F=\ell$,  
and we denote a representative of the equivalence class $[C_i]$ as
\begin{align}
  (\bse^{(i)}_1\cdots\bse^{(i)}_{\ell_i})\,,
\end{align}
with $t(\bse^{(i)}_{a})=s(\bse^{(i)}_{a+1})$ and $t(\bse^{(i)}_{\ell_i})=s(\bse^{(i)}_{1})$.
We also assume that each cycle $[C_i]$ ($i=1,\cdots,F$) has $b(C_i)$ bumps.
Then, the term in the expansion of $\frac{1}{\ell!}\left(\bar\eta B_u\eta\right)^{\ell}$ corresponding to the fermionic cycle $[\Psi]$ can be evaluated as 
\begin{align}
  \prod_{i=1}^F 
  u^{b(C_i)}
  &\left(\eta_{\bse^{(i)}_{1}}\bar{\eta}_{{\bse}^{(i)}_2}\right)
  \cdots
  \left(\eta_{\bse^{(i)}_{\ell_i}}\bar{\eta}_{{\bse}^{(i)}_1}\right) 
  =
  (-1)^{F+\ell} u^{b(\Psi)} \prod_{i=1}^F 
  \left(\bar{\eta}_{\bse^{(i)}_{1}}{\eta}_{{\bse}^{(i)}_1}\right)
  \cdots
  \left(\bar\eta_{\bse^{(i)}_{\ell_i}}{\eta}_{{\bse}^{(i)}_{\ell_i}}\right) \,,
\end{align}
where $b(\Psi)\equiv b(C_1)+\cdots+b(C_F)$ is the total number of the bumps in the fermionic cycle $[\Psi]$.
The term $\frac{1}{(2n_E-\ell)!}(\bar{\boldsymbol{\eta}}\boldsymbol{\eta})^{2n_E-\ell}$ in \eqref{eq:zeta_fermion tmp1} supplements the remaining Grassmann variables to form the total $2n_E$ Grassmann variables.
As a result, we can further rewrite \eqref{eq:zeta_fermion tmp1} as 
\begin{align}
  \zeta_G(q,u)^{-1}
  &= 
  \sum_{[\Psi]} (-1)^{F+\ell(\Psi)} u^{b(\Psi)}
  (-q)^{\ell(\Psi)}
  \int \prod_{\bse\in E_D}d\eta_\bse d\bar\eta_\bse\,
  \prod_{\bse\in E_D}\bar\eta_{\bse}{\eta}_{{\bse}} \nn \\
%  &= \sum_{[C_P]} (-1)^{F+\ell(C_p)} u^{b(C_P)}
%  (-q)^{\ell(C_P)} \nn \\
  &= 1+\sum_{[\Psi]} \mu(\Psi) u^{b(\Psi)}q^{\ell(\Psi)},
\end{align}
where we have used \eqref{eq:fermion measure} %to derive the second line 
and defined the cycle M\"obius function $\mu(C)$ by
\be
\mu(C) = \begin{cases}
  0 & \text{if the same directed edge is included somewhere in $C$}\\
  (-1)^{F} & \text{if $C$ contains $F$ distinct primitive cycles}
\end{cases}\, .
\ee
%in the third line. 
%Now let us introduce the ``prime'' cycles on the graph.
%A cycle $\hat{C}$ is called prime if it is a cycle
%that do not pass through the same edge twice\footnote{
%  In the context of graph zeta function, the prime cycle here is sometimes used in the same sense as the primitive cycle. The prime cycle is to be understood as a term only here.
%}.
%Any cycle on the graph can be constructed by concatenating the prime cycles like $C=\hat{C}_1\hat{C}_2\cdots\hat{C}_n$.
%Let $C_P$ be a product of the primitive cycles on the graph.
%For the product of the primitive cycles, we can define 
%Then, we find that the fermionic partition function is expressed as
%\be
%\zeta_G(q,u)^{-1} = 1+\sum_{[C_P]}
%\mu(C_P) u^{b(C_P)}q^{\ell(C_P)},
%\ee
%where $[C_P]$ runs over all products of the equivalence class of the primitive cycles on the graph,
%and $b(C_P)$ and $\ell(C_P)$ are the total number of the bump count and the length of the cycles including in $C_P$, respectively.
%Note that the sum can be taken over the product of all cycles, since non-primitive cycles are automatically excluded by the cycle M\"obius function.
Note that we do not need to restrict the summation of the last line only to the fermionic cycle 
but can take over all sets of the equivalence classes of the primitive cycles on the graph 
since the cycle M\"obius function limits terms to only the product of fermionic cycles.

The cycle M\"obius function does not allow the overlapping of directed edges due to the exclusion principle, and its signature makes it an alternating sum according to the number of the fermionic cycles, like the Witten index.
We then denote the fermionic cycles of length $\ell$ as $\Psi_{\bse_1\bse_2\cdots\bse_\ell}$, which is also a primitive cycle by definition.
In the sense of original fermions on the graph,
the fermionic cycle is a composite operator
(ordered product) of the fermions on the edges:
\be
\Psi_{\bse_1\bse_2\cdots\bse_\ell}
=\psi_{\bse_1}\psi_{\bse_2}\cdots\psi_{\bse_\ell},
\ee
where we have defined $\psi_{\bar{e}}\equiv \tilde{\psi}_{e}$.

\begin{table}[h]
  \begin{center}
  \begin{tabular}{|c|c|c|c|}
    \hline
    length & bumps & fermionic cycles & $F$\\
    \hline\hline
    2 & 2 & $\Psi_{1\bar{1}}$, $\Psi_{2\bar{2}}$, $\Psi_{3\bar{3}}$ & 1\\
    \hline
    3 & 0 & $\Psi_{123}$, $\Psi_{\bar{1}\bar{2}\bar{3}}$ & 1\\
    \hline
    \multirow{2}{*}{4} & 2 & $\Psi_{12\bar{2}\bar{1}}$, $\Psi_{23\bar{3}\bar{2}}$, $\Psi_{31\bar{1}\bar{3}}$ & 1\\
    \cline{2-4}
    & 4 & $\Psi_{1\bar{1}}\Psi_{2\bar{2}}$, $\Psi_{2\bar{2}}\Psi_{3\bar{3}}$, $\Psi_{3\bar{3}}\Psi_{1\bar{1}}$ & 2\\
    \hline
    \multirow{4}{*}{6} & 0 & $\Psi_{123}\Psi_{\bar{1}\bar{2}\bar{3}}$ & 2\\
    \cline{2-4}
    & 2 & $\Psi_{123\bar{3}\bar{2}\bar{1}}$, $\Psi_{231\bar{1}\bar{3}\bar{2}}$, $\Psi_{312\bar{2}\bar{1}\bar{3}}$ & 1\\
    \cline{2-4}
    & 4 & $\Psi_{1\bar{1}}\Psi_{23\bar{3}\bar{2}}$, $\Psi_{2\bar{2}}\Psi_{31\bar{1}\bar{3}}$, $\Psi_{3\bar{3}}\Psi_{12\bar{2}\bar{1}}$ & 2\\
    \cline{2-4}
    & 6 & $\Psi_{1\bar{1}}\Psi_{2\bar{2}}\Psi_{3\bar{3}}$ & 3\\
    \hline
  \end{tabular}
  \end{center} 
  \caption{Fermionic cycles appearing in the series expansion of the inverse of the Bartholdi zeta function of the cycle graph $C_3$.}
    \label{table:C3}
  \end{table}

\begin{figure}[h]
  \begin{center}
  \includegraphics[scale=0.6]{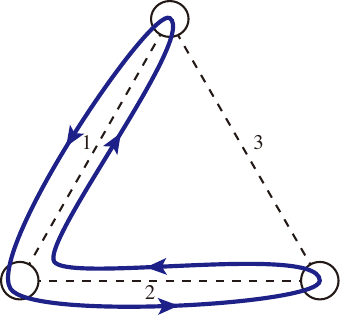}
  \end{center}
  \caption{An example of the fermionic cycle $\Psi_{12\bar{2}\bar{1}}$ on the cycle graph $C_3$. This cycle is a single primitive cycle with length four, two bumps and $F=1$.}
  \label{fermionic cycle in C3}
\end{figure}

As a concrete example, %let us first see the cycle graph $C_3$ with three vertices and three edges,
let us again consider the cycle graph $C_3$ 
%which is 
depicted in Fig.~\ref{cycle graph C3}.
The series expansion of the inverse of the Bartholdi zeta function of the cycle graph $C_3$ is given by \eqref{C3 inverse Bartholdi}.
%\be
%\zeta_{C_3}^{-1}(q,u)=1-3 u^2 q^2 -2 q^3-\left(3 u^2-3 u^4\right)q^4 +\left(1-3u^2+3 u^4-u^6\right)q^6 \, .
%\ee
Each term in this expansion can be read off from the fermionic cycles of the graph.
For example, the fermionic cycles of length $2$ are 
$[e_1\bar{e}_1]$,  
$[e_2\bar{e}_2]$ and 
$[e_3\bar{e}_3]$, 
which all include one primitive cycle and have two bumps%
\footnote{
  For example, for the cycle $(e_1\bar{e}_1)$, both of the edges $e_1$ and $\bar{e}_1$ are counted as bumps since the next edge is the inverse of the previous edge.
  Therefore $b((e_1\bar{e}_1))=2$. 
}. 
It corresponds to the result that the coefficient of $q^2$ in the expansion is $-3u^2$.
The other coefficients of this expansion are also reproduced from the list of the fermionic cycles of each length shown in Table \ref{table:C3}. An example of the fermionic cycle is shown in Fig.~\ref{fermionic cycle in C3}.

\begin{table}[h]
  \begin{center}
  \begin{tabular}{|c|c|c|}
    \hline
    length & fermionic cycles & $F$\\
    \hline\hline
    3 & $\Psi_{125}$, $\Psi_{\bar{5}\bar{2}\bar{1}}$, $\Psi_{\bar{5}34}$, $\Psi_{\bar{4}\bar{3}5}$ & 1\\
    \hline
    4 & $\Psi_{1234}$, $\Psi_{\bar{4}\bar{3}\bar{2}\bar{1}}$ & 1\\
    \hline
    6 & $\Psi_{125}\Psi_{\bar{5}\bar{2}\bar{1}}$, $\Psi_{\bar{5}34}\Psi_{\bar{4}\bar{3}5}$, $\Psi_{125}\Psi_{\bar{5}34}$, $\Psi_{\bar{4}\bar{3}5}\Psi_{\bar{5}\bar{2}\bar{1}}$ & 2\\
    \hline
    7 & $\Psi_{125}\Psi_{\bar{4}\bar{3}\bar{2}\bar{1}}$, $\Psi_{1234}\Psi_{\bar{5}\bar{2}\bar{1}}$, $\Psi_{\bar{5}34}\Psi_{\bar{4}\bar{3}\bar{2}\bar{1}}$, $\Psi_{1234}\Psi_{\bar{4}\bar{3}5}$ & 2\\
    \hline
    8 & $\Psi_{1234}\Psi_{\bar{4}\bar{3}\bar{2}\bar{1}}$ & 2\\
    \hline
    \multirow{2}{*}{10} & $\Psi_{125\bar{4}\bar{3}\bar{2}\bar{1}\bar{5}34}$, $\Psi_{\bar{4}\bar{3}51234\bar{5}\bar{2}\bar{1}}$, 
     & 1\\
    \cline{2-3}
    & $\Psi_{125}\Psi_{\bar{5}34}\Psi_{\bar{4}\bar{3}\bar{2}\bar{1}}$, 
    $\Psi_{1234}\Psi_{\bar{4}\bar{3}5}\Psi_{\bar{5}\bar{2}\bar{1}}$ & 3\\
    \hline
  \end{tabular}
  \end{center}
  \caption{Fermionic cycles appearing in the series expansion of the inverse of the Ihara zeta function of the double triangle graph DT.}
    \label{table:DT}
\end{table}

\begin{figure}[h]
  \begin{center}
  \subcaptionbox{$\Psi_{1234}\Psi_{\bar{5}\bar{2}\bar{1}}$}[.45\textwidth]{
  \includegraphics[scale=0.6]{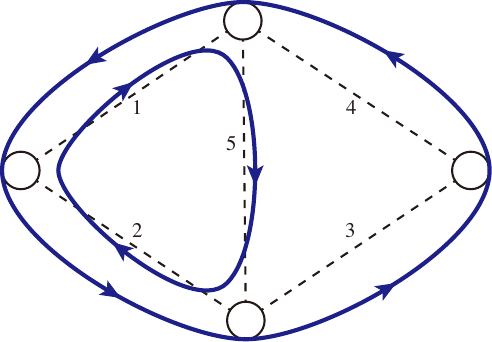}
  }
  \subcaptionbox{$\Psi_{\bar{4}\bar{3}51234\bar{5}\bar{2}\bar{1}}$}[.45\textwidth]{
  \includegraphics[scale=0.6]{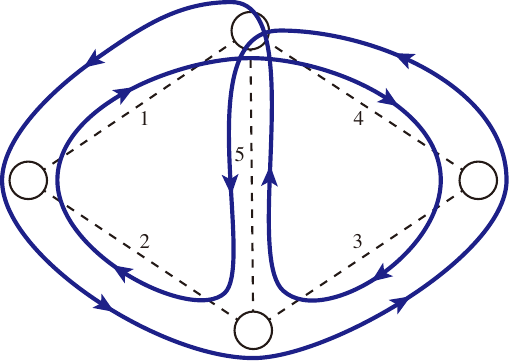}
  }
  \end{center}
  \caption{Examples of the fermionic cycles on the double triangle graph (DT).
  The left figure (a) shows the fermionic cycle $\Psi_{1234}\Psi_{\bar{5}\bar{2}\bar{1}}$
  with length 7, which is a product of two primitive cycles, then $F=2$.
  The right figure (b) shows the fermionic cycle $\Psi_{\bar{4}\bar{3}51234\bar{5}\bar{2}\bar{1}}$ with maximal length 10, which is a single primitive cycle of $F=1$.}
  \label{fermionic cycle in DT}
\end{figure}

As another example, %we consider the double triangle graph (DT), which
%has four vertices and five edges as depicted in Fig.~\ref{double triangle graph}.
let us consider DT depicted in Fig.~\ref{double triangle graph}.
%In this example, it is complicated to include bumps, 
%so we look only at the Ihara zeta function by setting $u=0$.
We again set $u=0$ to avoid unnecessary complications.
The series expansion of the inverse of the Ihara zeta function is given by \eqref{Inverse of Ihara for DT}. 
%The inverse of the Ihara zeta function of the double triangle graph has the series expansion in $q$, 
%\be
%\zeta_{\rm DT}(q)^{-1}=1-4q^3-2q^4+4q^6+4q^7+q^8-4q^{10}\,.
%\ee
As same as the previous example of the cycle graph $C_3$,
we see that the coefficients of this expansion are reproduced from the list of the fermionic cycles of each length shown in Table \ref{table:DT}. Two examples of the fermionic cycles are also shown in Fig.~\ref{fermionic cycle in DT}.

\section{Grid Graph}
\label{sec:grid graph}

In general, we need all data of the graph to evaluate the graph zeta function.  However, if the graph is a grid graph, that is, a graph consisting of periodic arrangement of a certain unit, the corresponding graph zeta function can be written explicitly by using only the information about the unit. 
In this section, we consider the model explained in the previous section on the grid graph.

\subsection{Covering graph}
\label{sec:covering graph}

The grid graph can be constructed by using the concept of 
%The idea of the construction of the grid graph comes from 
the covering (derived) graph
\cite{stark1996zeta,terras_2010,Clair2013TheIZ,Lenz2014TheIZ}. Let us now consider a finite group $G$ in addition to the digraph $\Gamma=(V,E)$ used so far.
A voltage assignment of $\Gamma$ by $G$ is a map $h_e: E \rightarrow G$, which assigns the group elements of $G$ on the edge $e\in E$.
The derived graph $\tilde{\Gamma}$ is constructed by the following way:
\begin{itemize}
\item The vertices of $\tilde{\Gamma}$ are the pairs $(v,g)$ of the vertex $v\in V$ of $\Gamma$ with the group element $g\in G$.
\item The edges of $\tilde{\Gamma}$ are the pairs $\langle(v,g),(v',h_e g)\rangle$ for each edge $e=\langle v,v'\rangle\in E$ of $\Gamma$.
\end{itemize}
Note also that there is a natural projection map $\pi: \tilde{\Gamma} \rightarrow \Gamma$ defined by $\pi(v,g)=v$.
Under this setup, 
it is known that the Bartholdi zeta function of the derived graph $\tilde{\Gamma}$ is expressed in terms of a product of
the Artin-Ihara $L$-function on the base graph $\Gamma$
\cite{Hashimoto1990ONZA,doi:10.1142/S0129167X92000370}
\be
\zeta_{\tilde{\Gamma}}(q,u) = \prod_{\rho} L_\Gamma(q,u;\rho)^{d_\rho},
\label{zeta by L}
\ee
with 
\begin{align}
  L_\Gamma(q,u;\rho) \equiv 
  &(1-(1-u)^2q^2)^{-(n_E-n_V)d_\rho} \nn \\
  & \times \det\left(
    I_{d_\rho n_V}
    -q \sum_{g\in G}  \rho(g)\otimes A_g 
    +q^2(1-u)I_{d_\rho}\otimes \left(D-(1-u)I_{n_V}\right) 
  \right)^{-1}\,,
\end{align}
where $\rho$ runs over the irreducible representations of the finite group $G$, 
$d_\rho$ is the multiplicity (dimension) of the representation $\rho$ 
and $A_g$ is a matrix of size $n_V$ whose elements are defined as $(A_g)_{v,v'}=1$ if $v$ and $v'$ are connected by an edge and $g\in G$ is assigned on the edge and $(A_g)_{v,v'}=0$ otherwise.

\begin{figure}[ht]
  \begin{center}
  \includegraphics[scale=0.4]{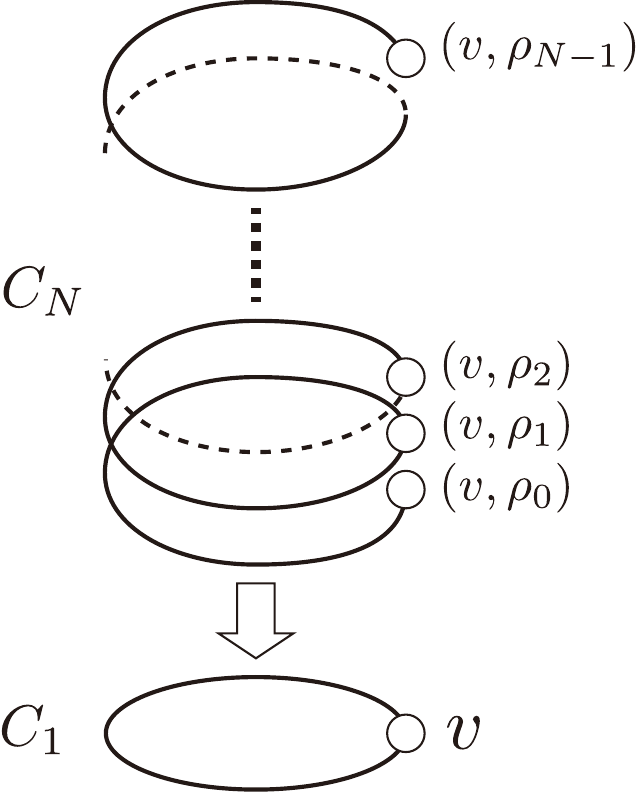}
  \end{center}
  \caption{The cycle graph $C_N$ as the covering graphs over $C_1$ graph. The voltage assignment is given by the representation of $\mathbb{Z}_N$. It is not shown in the figure, but the vertex at $(v,\rho_{N-1})$ is again connected to the vertex at $(v,\rho_0)$ owing to the periodic boundary condition.}
  \label{covering diagrams}
\end{figure}

The simplest example of the covering graphs is a cycle graph
$C_N$, whose graph zeta function can be constructed from
the $L$-function of the cycle graph $C_1$.
In this case, the finite group $G$ is the cyclic group $\mathbb{Z}_N$ and irreducible representations for the voltage assignment is given by powers of the $N$-th root of unity
\be
\rho_n \equiv \omega^n=e^{2\pi i n/N}
\qquad (n=0,\cdots,N-1).
\ee
The corresponding $L$-function of the cycle graph $C_1$ is given by
\be
L_{C_1}(q,u;\rho_n) = \frac{1}{1+q^2(1-u^2)-(\omega^n+\omega^{-n}) q}.
\label{eq:L-function C1}
\ee
and we can explicitly check \eqref{zeta by L} as
\be
\begin{split}
\zeta_{C_N}(q,u)
&=\det\left((1+q^2(1-u^2))I_N-q A_{C_N}\right)^{-1}\\
&= \prod_{n=0}^{N-1} L_{C_1}(q,u;\rho_n)
\, ,
\end{split}
\ee
where $A_{C_N}$ is the adjacency matrix of the cycle graph $C_N$. 

\subsection{Discrete Fourier analysis}

%Thus, 
%In the same way, we can construct the %complicated 
%graph zeta function of a grid graph 
%%from 
%as the product of $L$-functions of a 
%%simple 
%smaller 
%graph %in the fundamental domain 
%over the irreducible representations. 
%

\begin{figure}[ht]
  \begin{center}
  \subcaptionbox{$1\times 1$ grid graph}[.40\textwidth]{
  \includegraphics[scale=0.7]{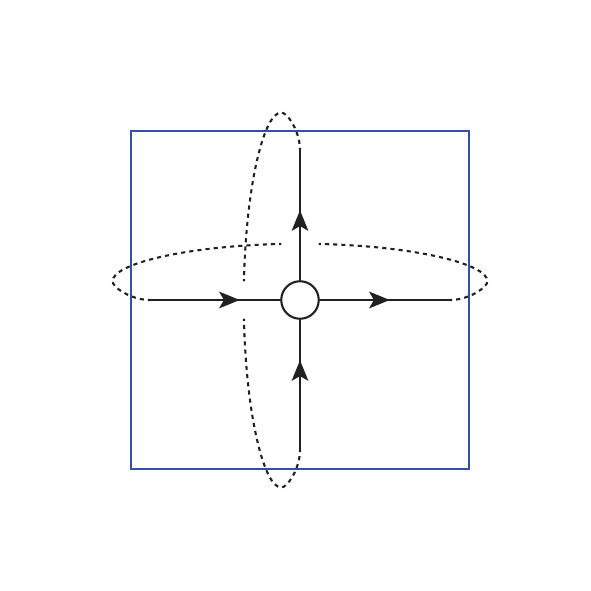}
  }
  \subcaptionbox{$4\times3$ grid graph}[.5\textwidth]{
  \includegraphics[scale=0.75]{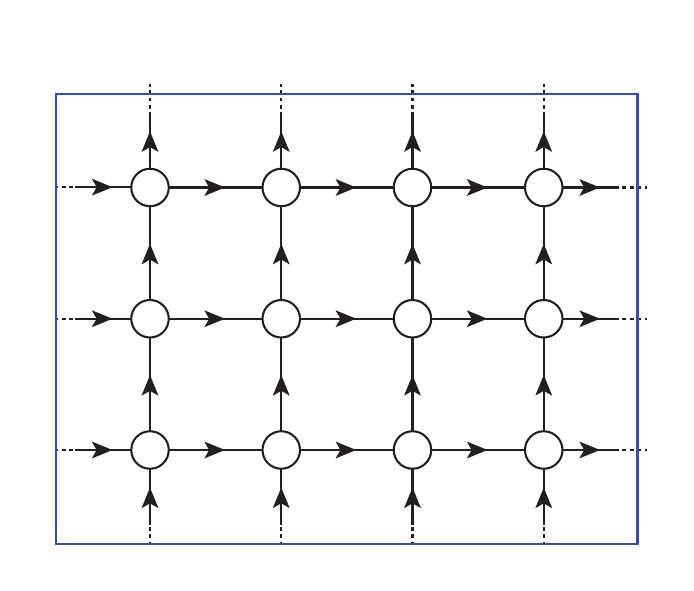}
  }
  \end{center}
  \caption{By using the discrete Fourier transformation, the square lattices are obtained from the covering space of the $1\times 1$ grid graph. The zeta function of the grid graph is expressed in terms of the $L$-function.}
  \label{grid diagrams}
\end{figure}

\begin{figure}[ht]
  \begin{center}
  \subcaptionbox{$1\times 1$ honeycomb grid graph}[.35\textwidth]{
  \includegraphics[scale=0.3]{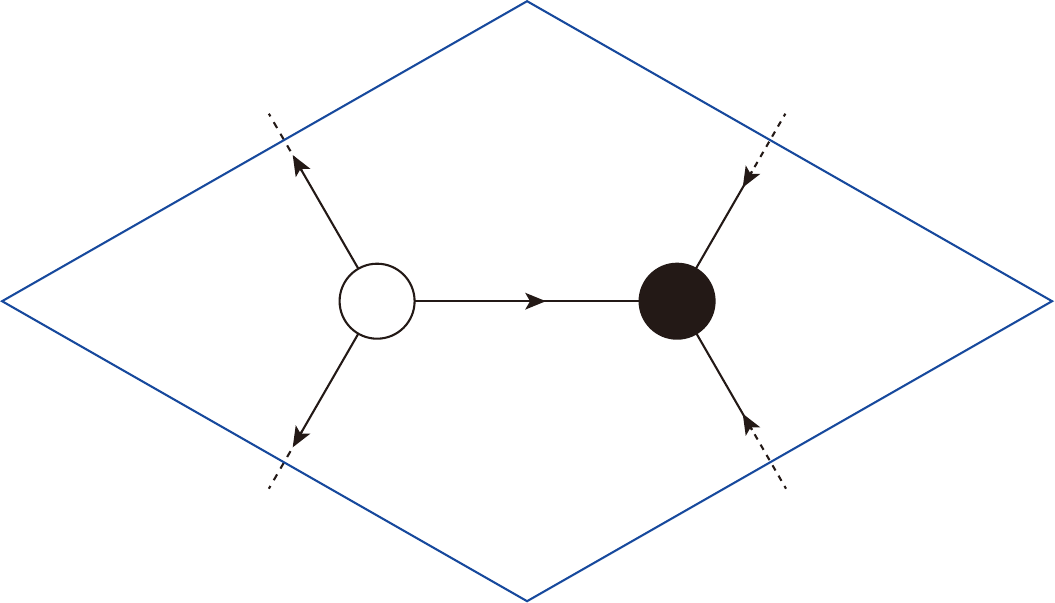}
  }
  \subcaptionbox{$4\times3$ honeycomb grid graph}[.55\textwidth]{
  \includegraphics[scale=0.32]{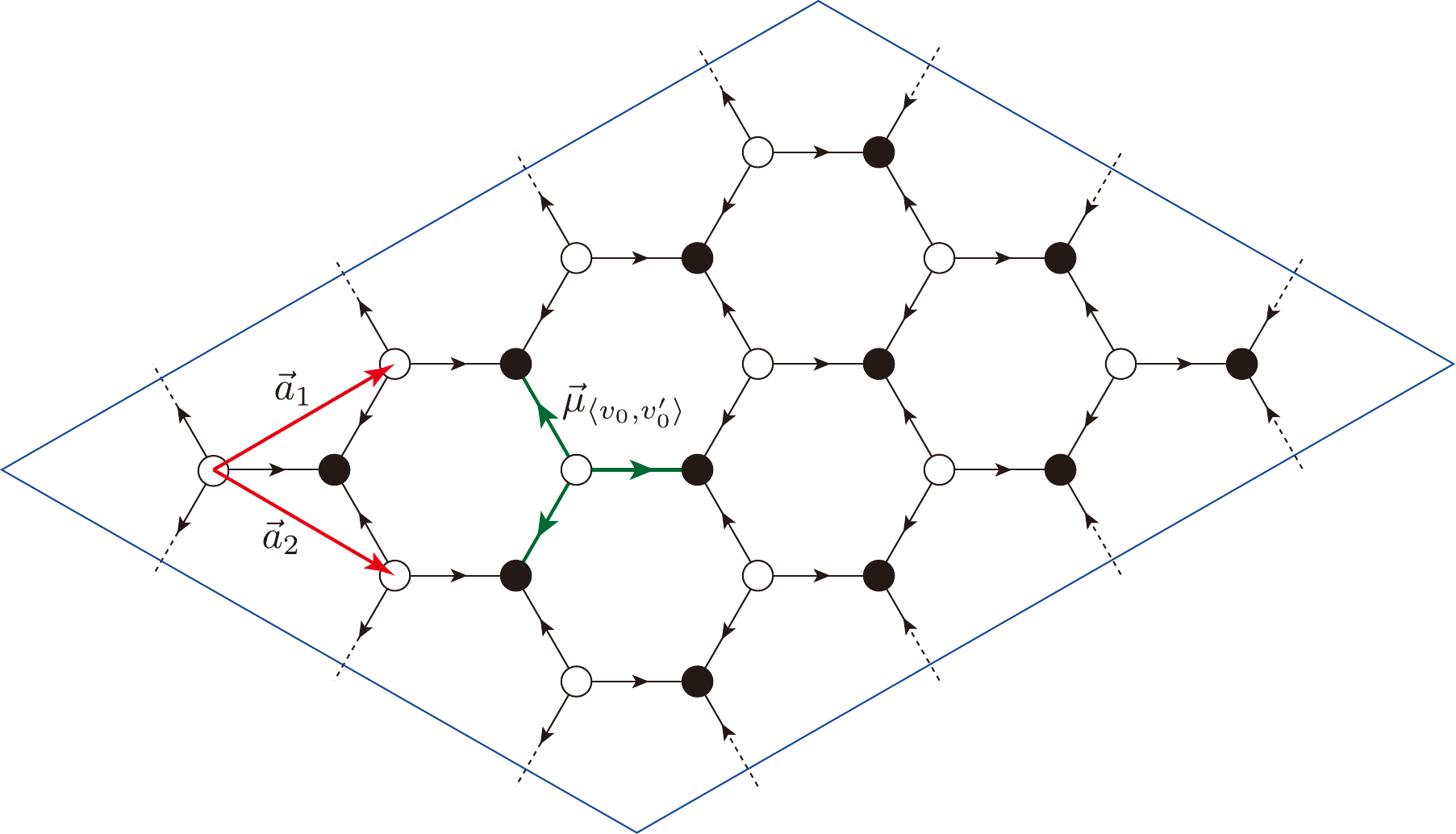}
  }
  \end{center}
  \caption{By using the discrete Fourier transformation, the honeycomb lattice on $T^2$ (b) is obtained from the covering space of the $1\times 1$ grid graph (a).
  The two red arrows represent the primitive basis vectors generating the honeycomb lattice.
  The three green arrows represent
  the direction vectors of the edges for the neighbors
  from white vertex to black vertex.}
  \label{honeycomb grid diagrams}
  \end{figure}

The cycle graph is not only the simplest example of the covering graph but also the simplest example of the grid graph, 
%which is obtained by repeating a unit graph periodically in several directions by connecting certain vertices in a unit graph to corresponding vertices in the next unit graph by reconnecting the corresponding edges in the unit graph as bridges between the two neighboring unit graphs 
which is constructed by reconnecting certain edges of a unit graph as {\it bridges} between two adjacent unit graphs
(see also Figs.~\ref{grid diagrams} and \ref{honeycomb grid diagrams} for examples of the two-dimensional gird). 
%In fact, we can regard the grid graph as a covering graph of a special kind.
This is a covering graph of a special kind. 
Let $\Gamma_0=(V_0,E_0)$ with $|E_0| \ge d$ ($d\in \mathbb{N}$) to be the unit graph and choose $d$ specific edges $e_1,\cdots, e_d\in E_0$.
In order to construct a covering graph, we consider a finite group $G=\mathbb{Z}_{N_1}\otimes \cdots \otimes \mathbb{Z}_{N_d}$
and assign $1\otimes \cdots \otimes \omega_i \otimes \cdots \otimes 1 \in G$ on the edge $e_i$ for $i=1,\cdots,d$, where $\omega_i$ is the $N_i$-th root of unity.
The yielding covering graph is nothing but a grid graph with $d$ independent periodicity. 
Therefore, we can evaluate the graph zeta function of the grid graph by applying the formula \eqref{zeta by L} to this setting.
In particular, since the group to construct the grid graph is an Abelian group and thus the irreducible representations are one-dimensional, the graph zeta function is simply a product of the $L$-functions.

In the following, we will show that the graph zeta function of the grid graph is evaluated in more familiar way for physicists by using the Fourier transformation. 
In this perspective, we can regard the $L$-function as a Fourier expansion of the graph zeta function.
We consider only the two-dimensional case ($d=2$) for simplicity,
but the generalization to higher dimensions is straightforward.
%a planar grid graph which is periodic only to two directions for simplicity. 
%The generalization to higher dimensions is straightforward.
%We will see that the description of the graph zeta function of the grid graph with respect to the $L$-function can be understood as the Fourier expansion of the graph zeta function.
%
%We next consider the graph zeta function of a general grid graph 
%and 
%We can extend this idea to construct a graph by repeatedly connecting a unit graph $(V_0,E_0)$ to more than one direction, 
%which yields a grid graph $(V,E)$. 
%we consider a planar grid graph $(V,E)$ on a two-dimensional torus $T^2$ for simplicity. 
%It is 

Although we do not need a coordinate space to define a graph, 
it is useful to draw the grid graph on a continuous torus $T^2$ for our purpose.
We call the unit cell of the grid graph on $T^2$ the fundamental domain 
and denote the directions of the primitive basis vectors $\vec{a}_1$ and $\vec{a}_2$. 
%We assume that the neighboring fundamental domains are connected 
%We also assume that the length of the edges (lattice spacing) is $a$.
Then, the coordinate of the torus is expressed by 
\be
\vec{x}= \vec{x}_0+n_1 \vec{a}_1+n_2 \vec{a}_2,
\quad (n_1=0,1,\cdots,N-1, \ n_2=0,1,\cdots,M-1)
\label{coordinate}
\ee
where $\vec{x}_0$ is the coordinate in the fundamental domain,
and it satisfies the periodic boundary conditions 
$\vec{x}\sim \vec{x}+N\vec{a}_1$ and $\vec{x}\sim \vec{x}+M\vec{a}_2$. 
%In sec.~\ref{sec:free fermion}, we have seen that it is useful to consider an appropriate field theory on the graph to evaluate the graph zeta function as the partition function of the theory. 
%In the case of the grid graph, the ``field'' (boson or fermion) can be written as $\phi(\vec{x}_v)$ at the positions $\vec{x}_v$ of the vertices $v$ on $T^2$.  
%By using the periodic structure, which are copies of $v_0$ in the fundamental domain, 
%then
%Since the graph has a periodic structure,
%we can relabel it in terms of a vertex $v_0 \in V_0$ in the fundamental domain 
%and integers $(n_1,n_2)$ as 
%\begin{equation}
%  \phi^v%(\vec{x}_v)
%  \equiv \phi(x_v) 
%  = \phi(\vec{x}_{v_0}+n_1\vec{a}_1+n_2\vec{a}_2)
%  \equiv \phi^{v_0}(\vec{n}) \,.
%  \label{relabel}
%\end{equation}
On the other hand, if we introduce the reciprocal lattice vectors $\vec{b}_1$ and $\vec{b}_2$ through the relation 
\be
\vec{a}_i\cdot \vec{b}_j = 2\pi \delta_{ij}\,,
\ee
the momentum is given by a vertex of the reciplocal lattice as
$\vec{k}=\frac{m_1}{N} \vec{b}_1+\frac{m_2}{M} \vec{b}_2$,
where $m_1=0,1,\cdots,N-1$ and $m_2=0,1,\cdots,M-1$ are momentum modes.

In Sec.~\ref{sec:free fermion}, we have seen that the graph zeta function can be evaluated as the partition function of a theory of complex bosonic fields $\phi^v$ on the graph as \eqref{zeta by boson}. 
Since the field $\phi^v$ has the periodicity to both of the directions $\vec{a}_1$ and $\vec{a}_2$, it has the discrete Fourier expansion
\be
\phi^v
= \frac{1}{\sqrt{NM}}\sum_{\vec{k}}
\hat{\phi}_{v_0}(\vec{k})
e^{i\vec{k}\cdot \vec{x}_v} \,,
%= \frac{1}{\sqrt{NM}}\sum_{m_1=1}^N \sum_{m_2=1}^M 
%\hat{\phi}_{v^0}(\vec{k})
%e^{i\vec{k}\cdot \vec{x}_{v^0}}
%e^{2\pi i (\frac{n_1m_1}{N}+\frac{n_2m_2}{M}) }\,.
\ee
where $\vec{x}_v=\vec{x}_{v_0}+n_1\vec{a}_1+n_2\vec{a}_2$ is the positions of the vertices $v$
corresponding to $v_0$ in the fundamental domain,
and the sum is taken over the momentum mode $\vec{m}$.
Note here that the Fourier mode $\hat{\phi}_{v_0}(\vec{k})$ is labeled by $v_0$
to specify which vertex in the fundamental domain corresponds to.
Substituting this Fourier expansion into the partition function \eqref{zeta by boson}, %with $\beta=2\pi$, 
%we can block diagonalize $\Delta_{q,u}$ in the momentum space and evaluate the Bartholdi zeta function as 
we can evaluate the partition function as a path integral over the momentum space
and obtain the Bartholdi zeta function of the grid graph as
\begin{multline}
%\begin{align}
  \zeta_\Gamma(q,u)
%  &= \left(1-(1-u)^2q^2\right)^{-(n_{E_0}-n_{V_0})NM}
%  \int \prod_{v\in V}d\phi^v d\bar\phi^v\, e^{-2\pi S_B(q,u)} \nn \\
%  &= \left(1-(1-u)^2q^2\right)^{-(n_{E_0}-n_{V_0})NM}
%  \int \prod_{v_0\in V_0} \prod_{\vec{k}} 
%  d \hat\phi_{v_0}(\vec{k}) d \bar{\hat\phi}_{v_0}(\vec{k}) \nn \\
%  &\qquad\qquad 
%  \prod_{\vec{k}}
%  e^{-2\pi  \sum_{v_0,v'_0\in V_0}
%  \bar{\hat\phi}_{v_0}(\vec{k}) \hat\phi_{v'_0}(\vec{k})
%  \left(
%    \delta_{v_0,v'_0} 
%    - q (\hat{A}_{\Gamma_0}({\vec{m}}))_{v_0 v'_0} 
%    + q^2(1-u)\left(\deg{v_0}-(1-u)\right) \delta_{v_0,v'_0}
%  \right)}\nn \\
%  & 
%  \mathscale{0.85}{
  =
  \prod_{m_1=0}^{N-1}\prod_{m_2=0}^{M-1}
  \left(1-(1-u)^2q^2\right)^{-(n_{E_0}-n_{V_0})}\\
  \times \det\left(
    I_{n_{V_0}} - q \hat{A}_{\Gamma_0}(\vec{m}) + q^2(1-u)\left(D_0-(1-u)I_{n_{V_0}}\right)
  \right)^{-1}
%  }
  \label{grid zeta}
%\end{align}
\end{multline}
where $n_{V_0}=|V_0|$, $n_{E_0}=|E_0|$, 
$D_0$ is the degree matrix of $\Gamma_0$
and $\hat{A}_{\Gamma_0}(\vec{m})$ is the adjacency matrix of size $n_{V_0}$ in the momentum space 
whose elements are defined by %as a summation over the edges and the vertices in the fundamental domain $E_0$ as
\begin{align}
  (\hat{A}_{\Gamma_0}(\vec{m}))_{v_0 v'_0} \equiv 
  \sum_{e \in E_0}
  (\delta_{v_0,s(e)}\delta_{v'_0,t(e)}
  +\delta_{v_0,t(e)}\delta_{v'_0,s(e)})
   \exp\left\{i \vec{k}\cdot \vec{\mu}_{\langle v_0,v_0'\rangle}\right\} 
\end{align}
with the direction vectors of the edges
$\vec{\mu}_{\langle v_0,v_0'\rangle}\equiv \vec{x}_{v_0'}-\vec{x}_{v_0}$. 
This reproduces the decomposition \eqref{zeta by L} of the Bartholdi zeta function by the $L$-function on the covering graph. 
Therefore, $L$-function can be understood as the Fourier expansion of the graph zeta function of the grid graph as announced.

For example, let us consider the familiar square lattice with $N\times M$ grid on the torus
(see Fig.~\ref{grid diagrams}).
The primitive basis vectors are
\be
\vec{a}_1=(a,0),\quad \vec{a}_2=(0,a)\, ,
\ee
and the corresponding reciprocal lattice vectors are
\be
\vec{b}_1=\left(\frac{2\pi}{a},0\right),
\quad \vec{b}_2=\left(0,\frac{2\pi}{a}\right)\, ,
\ee
where we have assumed that the lattice spacing is $a$. 
The $1\times 1$ grid graph in the unit cell
consists of a single vertex and four edges.
Since the degree of the vertex on the grid graph is 4 
and 
the direction vectors of the edges for the neighbors
are given by
$\vec{\mu}_{\langle v_0,v_0'\rangle}=(a,0),\ (-a,0),\ (0,a),\ (0,-a)$,
%where the opposite directed edges are also distinguished.
%we find the following factorization identity
%by comparing the determinants between the coordinate and
%momentum spaces
we obtain 
\be
\begin{split}
  \zeta_{\rm SQ}(q,u)^{-1} %&=(1-q^2(1-u)^2)^{NM}\det\left((1+(1-u)(3+u)q^2 )I_{NM}-q A_{\rm SQ}\right)\\
&=\prod_{m_1=0}^{N-1}\prod_{m_2=0}^{M-1}
(1-q^2(1-u)^2)
\left(1+(1-u)(3+u)q^2-q \hat{A}_{\rm SQ}(\vec{m})\right)\, ,
\end{split}
\label{SQ grid zeta}
\ee
where
%$A_{\rm SQ}$ is the $NM\times NM$ adjacency matrix of the square lattice and 
%$\hat{A}_{\rm SQ}(\vec{m})$ is the adjacency matrix in the momentum space
\be
\hat{A}_{\rm SQ}(\vec{m})
=\omega_1^{m_1}+\omega_1^{-m_1}+\omega_2^{m_2}+\omega_2^{-m_2}
\, ,
\ee
with $\omega_1\equiv e^{2\pi i/N}$ and $\omega_2\equiv e^{2\pi i/M}$.
%are the $N$-th and $M$-th roots of unity, respectively.
%This expression implies that a general $N\times M$ grid 
%graph can be regarded as a covering space of the $1\times 1$ grid graph, as a sort of the $L$-function on the graph.

The second example is the honeycomb lattice depicted in Fig.~\ref{honeycomb grid diagrams}.
The two-dimensional primitive basis vectors are spanned by
\be
\vec{a}_1=\left(\frac{3a}{2},\frac{\sqrt{3}a}{2}\right),\quad \vec{a}_2=\left(\frac{3a}{2},-\frac{\sqrt{3}a}{2}\right)\, ,
\ee
where $a$ is the distance between the neighboring vertices.
The reciprocal lattice vectors are
\be
\vec{b}_1=\left(\frac{2\pi}{3a},\frac{2\pi}{\sqrt{3}a}\right),
\quad \vec{b}_2=\left(\frac{2\pi}{3a},-\frac{2\pi}{\sqrt{3}a}\right)\, .
\ee
The $1\times 1$ honeycomb grid graph in the unit cell
consists of two vertices and three edges.
The direction vectors of the edges for the neighbors
are given by
\be
\vec{\mu}_{\langle v_0,v_0'\rangle}=(a,0),\ 
\left(-\frac{a}{2},\frac{\sqrt{3}a}{2}\right),\ 
\left(-\frac{a}{2},-\frac{\sqrt{3}a}{2}\right)\, ,
\ee
and their opposites.
Since the phases coming from
the norms with the momentum vectors are given by
\be
e^{i\vec{k}\cdot \vec{\mu}_{\langle v_0,v_0'\rangle}}
=\omega_1^{\frac{m_1}{3}}\omega_2^\frac{m_2}{3}\, ,\quad
\omega_1^\frac{m_1}{3}\omega_2^{-\frac{2m_2}{3}}\, ,\quad
\omega_1^{-\frac{2m_1}{3}}\omega_2^\frac{m_2}{3}\, ,
\ee
and their inverses,
the adjacency matrix in the momentum space becomes 
\begin{multline}
\hat{A}_{\rm HC}(\vec{m})\\=\begin{pmatrix}
  0 & \omega_1^\frac{m_1}{3}\omega_2^\frac{m_2}{3}
  +\omega_1^\frac{m_1}{3}\omega_2^{-\frac{2m_2}{3}}
  +\omega_1^{-\frac{2m_1}{3}}\omega_2^\frac{m_2}{3}\\
  \omega_1^{-\frac{m_1}{3}}\omega_2^{-\frac{m_2}{3}}
  +\omega_1^{-\frac{m_1}{3}}\omega_2^{\frac{2m_2}{3}}
  +\omega_1^{\frac{2m_1}{3}}\omega_2^{-\frac{m_2}{3}} & 0
\end{pmatrix}\, .
\end{multline}
We then find 
\be
\begin{split}
  \zeta_{\rm HC}(q,u)^{-1}
  %&=(1-q^2(1-u)^2)^{NM}\det\left((1+(1-u)(2+u)q^2)I_{2NM}-q A_{\rm HC}\right)\\
  &=\prod_{m_1=0}^{N-1}
  \prod_{m_2=0}^{M-1}
  (1-q^2(1-u)^2)\det\left((1+(1-u)(2+u)q^2)I_2-q\hat{A}_{\rm HC}(\vec{m})\right)\, .
  \label{eq:zeta HC}
\end{split}  
\ee
More concretely, if we set $u=0$, $N=4$ and $M=3$
($n_V=24$ and $n_E=36$),
the fermion partition function (the inverse of the Ihara zeta function) for the $4\times 3$ honeycomb lattice has
the factorized form
\be
\begin{split}
\zeta_{\rm HC}(q)^{-1}&=(1-q^2)^{12}\prod_{m_1=0}^{3}\prod_{m_2=0}^{2}
\det\left((1+2q^2)I_2-q\hat{A}(\vec{m})\right)\\
&=1-32 q^6-78 q^8-240 q^{10}-80 q^{12}
+96 q^{14}+2487 q^{16}+{\cal O}\left(q^{18}\right),
\end{split}
\ee
which agrees with the explicit calculation of the Ihara zeta function
by using the $24\times 24$ adjacency matrix in the covering grid graph.
We also see that the series expansion in $q$ correctly counts 32 shortest length 6 cycles in Fig.~\ref{honeycomb grid diagrams} (b).

The advantage of the representation of the graph zeta function on the grid by using the product of the $L$-function is that the distribution of the poles of the graph zeta function can be studied in the continuum limit where the number of grids is very large.
The poles of the graph zeta function are the zeros of its inverse, namely the zeros of the partition function of the fermions, which are obtained as the poles of the $L$-functions in the product over the grids.

\begin{figure}[ht]
  \begin{center}
  \subcaptionbox{$100\times 100$ square lattice}[.45\textwidth]{
  \includegraphics[scale=0.8]{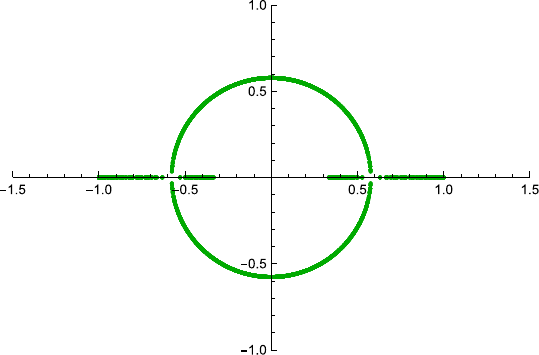}
  }
  \subcaptionbox{$100\times 100$ honeycomb lattice}[.45\textwidth]{
  \includegraphics[scale=0.8]{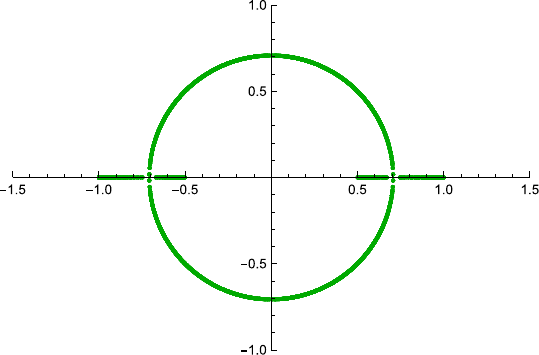}
  }
  \end{center}
  \caption{The poles of the Ihara zeta function (zeros of the partition function of the fermionic model) for the $100\times 100$ square lattice (a) and the $100\times 100$ honeycomb lattice (b) in the complex $q$-plane.}
  \label{zeros of the grids}
\end{figure}

In Fig.~\ref{zeros of the grids}, we show the positions of the poles of the Ihara zeta function (zeros of the partition function) for the $100\times 100$ square lattice and the $100\times 100$ honeycomb lattice in the complex $q$-plane.

As discussed in \cite{Matsuura:2024gdu} and summarized in the text book \cite{terras_2010},
the poles of the Ihara zeta function are distributed in the complex $q$-plane in the following way:
For $(p+1)$-regular graphs, where each vertex has the same degree $p+1$,
we can show that the real poles of the Ihara zeta function are located on
the line segments of $1/p \leq |q| \leq 1$ for $q\in \mathbb{R}$,
and others are located on a circle of $|q|=1/\sqrt{p}$.
The square and honeycomb lattices are the case of $p=3$ and $p=2$, respectively,
then the distribution of the poles of the Ihara zeta function of the square and honeycomb lattices in Fig.~\ref{zeros of the grids} is consistent with this fact.

Incidentally, a connected regular graph is called a Ramanujan graph if the eigenvalues $\lambda$ of the adjacency matrix satisfy $\lambda \leq 2\sqrt{p}$, except for the largest eigenvalue $|\lambda|=p+1$. The poles on the line segment on the real axis corresponds to the non-Ramanujan eigenvalues. So if there is no pole on the line segment except for the boundary $q=\pm 1, \pm 1/p$, the graph becomes Ramanujan and satisfies the Riemann hypothesis by redefining the parameter $q$ as $q=p^{-s}$,
namely the non-trivial poles of the Ihara zeta function are located only on the critical line ${\rm Re}\, s=1/2$.

Although there is not much difference in the distributions of poles of the original graph zeta function between square and honeycomb lattices, 
we will see this in the next section that they change drastically by deforming the zeta function by a certain parameter where the distribution of the poles has quite interesting physical meanings.

%%%
\subsection{Absence of the fermion doubling}
%
%We have given above a general prescription to construct the fermion partition function of our model on the grid graph as
%products over $L$-functions.
We now regard the inverse of the graph zeta function of the grid graph \eqref{zeta by L} as the partition function of the fermionic model with the action \eqref{fermion action}.  
%defined in sec.~\ref{sec:fermion theory}. 
Unlike general random graphs, 
%the consideration of periodic lattices  
the periodic structure of the grid graph allows us to introduce momentum and its dispersion relations. 
%since translational symmetry emerges.
%When translational symmetry appears in fermions on the periodic lattice, 
%At the same time, 
Inevitably, we are concerned with the fermion doubling problem,
since translational symmetry is one of the conditions for the Nielsen-Ninomiya theorem \cite{Nielsen:1980rz,Nielsen:1981xu}.
%Here we would like to discuss the relation to the fermion doubling problem.

From the discussion in the previous subsection, 
we can generally rewrite the action \eqref{fermion action} in the momentum space as 
\be
  S = \sum_{\vec{k}}
  \hat{\bar{\Psi}}(\vec{k})
  \left(\slashed{D}_{\vec{k}} + {\cal M}\right)
  \hat\Psi(\vec{k})\,, 
\ee
where $\hat\Psi(\vec{k})$, $\hat{\bar{\Psi}}(\vec{k})$ and $\slashed{D}_{\vec{k}}$ are the Fourier transformations of the fermions $\Psi$ and $\bar{\Psi}$ and the operator $\slashed{D}$ on the grid graph, respectively.
%We can read off the spectrum of the fermion from 
Since the ``propagator'' of the matrix $\slashed{D}_{\vec{k}}+{\cal M}$ can be evaluated as 
\begin{align}
  \left(\slashed{D}_{\vec{k}}+{\cal M}\right)^{-1}
  &= \det\left(\slashed{D}_{\vec{k}}+{\cal M}\right)^{-1}
  \left(\slashed{D}_{\vec{k}}+{\cal M}\right)^+\,,
\end{align}
where $\left(\slashed{D}_k+{\cal M}\right)^+$ stands for the classical adjoint of $\slashed{D}_k+{\cal M}$, 
the physical modes in the continuum limit can be read off by expanding $\det\left(\slashed{D}_k+{\cal M}\right)$ by $\vec{k}$ around the minimum. 
Recalling that $\det\left(\slashed{D}_{\vec{k}}+{\cal M}\right)$ is nothing but the inverse of the $L$-function $L_\Gamma(q,u;\vec{k})^{-1}$ of the base graph $\Gamma$,
it is sufficient to consider the expansion of the $L$-function around the minimum.

As a typical example, let us consider the square lattice. 
%In order to see if there exists a species doubler, we have only to consider the $1$d lattice whose L-function is given by \eqref{eq:L-function C1}. 
In this case, by writing $\omega_i \sim e^{ik_i a}$ ($i=1,2$), 
the $L$-function is expressed from (\ref{SQ grid zeta}) as 
\begin{align}
  L_{\rm SQ}(q,u;\vec{k})^{-1} = 1+q^2(1-u^2)-2q (\cos k_1a+\cos k_2a)\,,
\end{align}
which has minimum only at $k=0$ for $q>0$ in the continuum limit. 
%and $k=\frac{\pi}{a}$ when $q<0$. 
The generalization to the (hyper)cubic lattice is straightforward.
So we can conclude that the fermions in the model on the square lattice do not have any species doubler.

We can also apply the same analysis to the honeycomb lattice. 
In this case, from \eqref{eq:zeta HC}, 
the $L$-function can be written as 
\begin{align}
  L_{{\rm HC}}(q,u;\vec{k})^{-1}
  = 
  &(1-q^2(1-u)^2)
  \left(1+(1-u)(2+u)q^2\right)^2 \nn \\
  &-q^2
  \left(3+2\cos (k_1a) 
  + 2 \cos (k_2 a) 
  + 2 \cos ((k_1+k_2)a) \right)\,.
\end{align}
%where HC$_{1,1}$ expresses the base graph of the honeycomb grid depicted in Fig.~\ref{honeycomb grid diagrams} (a) and we have expressed $\omega_i = e^{ik_i a}$ ($i=1,2$). 
Again, since the minimum of the $L$-function is only at $k_1=k_2=0$, there is no fermion species doubler. 

%We can understand the absence of the fermion doubler %also can be understood from another perspective. 
%from the structure of the fermions on the graph. 
%%To see this more explicitly, %let us consider a change 
%In fact, the matrix $\slashed{D}$ is constructed from 
%the (deformed) incidence matrix which maps from the vertices $V$ to the edges $E$. 
%It is a discrete analog of the exterior derivative of the differential form, which maps from the 0-form to 1-form. 
%Thus, the fermions have essentially the same structure of the K\"ahler-Dirac fermion on the two-dimensional lattice \cite{Banks:1982iq} or equivalently the staggered fermion \cite{Bodwin:1987ah}, 
%which is known that there is no fermion doubler.
%
%For example, if we consider the two-dimensional grid graph, we can combine the fermions on the vertices and edges into multicomponent fermions associated a bipartite structure on the grid graph as shown in Fig.~\ref{staggered fermion}. The bipartite structure can be regarded effectively as doubling the lattice spacing $a$ to $2a$, as well as the discussion of the staggered fermion. 
%This is the essential reason why our model avoids the fermion doubling problem. 

\begin{figure}[t]
  \begin{center}
  \includegraphics[scale=0.7]{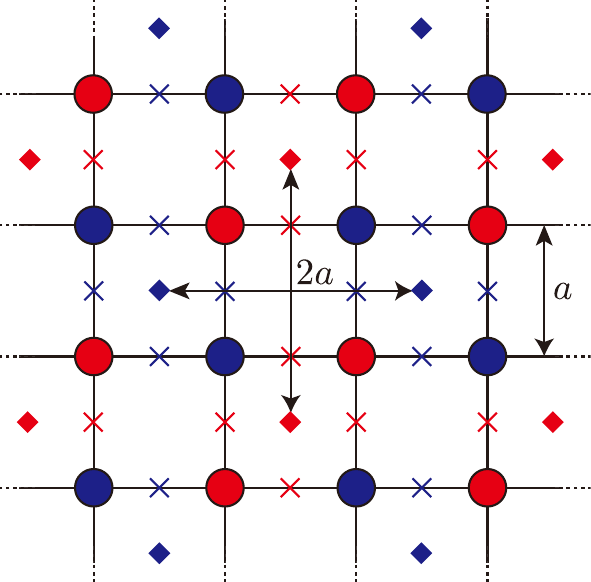}
  \end{center}
  \caption{A schematic explanation of the absence of the fermion doubler as the staggered-like fermion on the two-dimensional grid graph (square lattice).
  We can combine the fermion on the vertices and edges into multicomponent fermions by assigning a bipartite structure on the grid graph. Two different colors (red and blue) represent the components of two different types of fermions.
  These fermions lie on the extended grid with twice of the original lattice spacing $a$.}
  \label{staggered fermion}
 \end{figure}

We can understand the absence of the fermion doubler from the structure of the fermions on the graph. 
The point is that the kinetic term of the model is defined as hopping terms between the vertices and edges of the graph as the K\"ahler-Dirac fermion on the two-dimensional lattice \cite{Banks:1982iq} 
%the matrix $\slashed{D}$ is constructed from 
%the (deformed) incidence matrix which maps from the vertices $V$ to the edges $E$ 
and 
the fermions on the vertices and edges are combined into multicomponent fermions associated with a bipartite structure on the grid graphs 
(see Fig.~\ref{staggered fermion} for the example of the square lattice).
The bipartite structure can be regarded effectively as doubling the lattice spacing $a$ to $2a$, as well as the discussion of the staggered fermion \cite{Bodwin:1987ah}. 
This is the essential reason why our model avoids the fermion doubling problem.

%Thus, the fermions have essentially the same structure of the K\"ahler-Dirac fermion on the two-dimensional lattice \cite{Banks:1982iq} or equivalently the staggered fermion \cite{Bodwin:1987ah}, 
%which is known that there is no fermion doubler.

%For example, if we consider the two-dimensional grid graph, we can combine the fermions on the vertices and edges into multicomponent fermions associated a bipartite structure on the grid graph as shown in Fig.~\ref{staggered fermion}. 

From another perspective, we also would like to point out similarities between our model and supersymmetric lattice gauge theories. %\cite{Sugino:2003yb, Sugino:2004qd,Kikukawa:2008xw,Misumi:2013maa}
%and its generalization to the graph \cite{Matsuura:2014kha,Matsuura:2014nga,Kamata:2016xmu,Ohta:2021jze}.
The Sugino model \cite{Sugino:2003yb, Sugino:2004qd}, which is two-dimensional supersymmetric Yang-Mills theory on the square lattice, is generalized to the graph \cite{Matsuura:2014kha,Matsuura:2014nga,Kamata:2016xmu,Ohta:2021jze}, where the gauge field $A_e$ is defined on the edge (link) $e$, 
the scalar fields $\phi^v$ and $\bar{\phi}^v$ are defined on the vertex (site) $v$, 
and the fermions $\eta^v$, $\lambda^e$ and $\chi^f$ are defined on the vertex $v$, edge $e$ and face (plaquette) $f$, respectively.
Setting the gauge group to $U(1)$, the supersymmetric transformation of the fields is given by
\be
\begin{array}{lcl}
Q \phi^v = 0, && \\
Q \bar{\phi}^v = 2\eta^v, && Q \eta^v = 0,\\
Q A^e = \lambda^e, && Q\lambda^e =-{L^e}_v\phi^v,\\
Q Y^f = 0, && Q\chi^f = Y^f\, ,
\end{array}
\label{SUSY transformation}
\ee
where
$L$ is the incidence matrix and 
$Y^f$ is the auxiliary field associated the face $f$ of the graph, needed for the off-shell formulation of the supersymmetry. Note that the above single supercharge $Q$ behaves as a scalar filed and survives even on the discrete space-time (graph).
% and moment map constraints $\mu^f$.
%The fermionic fields $(\eta^v,\lambda^e,\chi^f)$
%are exists on the vertices, edges and faces of the graph, respectively.
Using this transformation, the supersymmetric action can be written in the $Q$-exact form
\be
S_{\rm SUSY} = Q\Xi \, ,
\ee
where
\be
\Xi = -\frac{1}{2g^2}\left\{\bar{\phi}_v{L^v}_e\lambda^e
+\chi_f(Y^f-2\mu^f)
\right\}\, .
\label{eq:Xi}
\ee
where $\mu^f$ is the moment map associated with the face $f$ of the graph.
%By choosing the moment map $\mu^f$ suitably,
%we can show that the 1-loop determinants of the bosonic and fermionic parts are cancelled with each other,
%namely the spectrum and dispersion relation of the bosons and fermions coincide.
In this formulation, the Wilson term naturally appears as a consequence of the dimensional reduction from higher dimensional theory with maximal supersymmetry, which prevent the fermion doublers from arising \cite{Sugino:2003yb, Sugino:2004qd,Kikukawa:2008xw,Misumi:2013maa}.
%Note that we can also understand it as a result of the supersymmetric correspondence of the dispersion relation between the bosons and fermions.
This also can be understood from the fact that the spectra and dispersion relations of bosons and fermions coincide with each other in supersymmetric theories.

On the other hand, by re-expressing the fermions of the present model to the Weyl basis %of the fermions 
\be
\Psi \to U\Psi =\left(\xi,\psi_R,\psi_L\right)^T\,,
%\bar{\Psi} \to \bar{\Psi}T^{-1} =\left(\xi,\bar{\psi}_R,\bar{\psi}_L \right)\,.
\ee
where $\psi_R\equiv \frac{1}{\sqrt{2}}(\psi+\psit)$ and $\psi_L\equiv-\frac{1}{\sqrt{2}}(\psi-\psit)$ are the right-handed and left-handed Weyl fermions, respectively, 
%In this chiral base, 
%the $\gamma_5$ matrix is diagonalized into $\gamma_5'=\diag(I_{n_V},I_{n_E},-I_{n_E})$
%and 
the Dirac operator can be rewritten as 
%\be
%\slashed{D} \to \slashed{D}'=U \slashed{D}U^{-1} 
%\, ,
%\ee
%where $\slashed{D}'$ is given by
\be
\slashed{D}'=
U \slashed{D}U^{-1}
= \frac{\alpha}{\sqrt{2}}
\begin{pmatrix}
  0 & L_{q,u}^T+\Lt_{q,u}^T & L_{q,u}^T-\Lt_{q,u}^T\\
  L_{q,u}+\Lt_{q,u} & 0 & 0\\
  -(L_{q,u}-\Lt_{q,u}) & 0 & 0
  \end{pmatrix}\, .
  \label{Dirac chiral}
\ee
Recalling $L_{q,u}$ and $\Lt_{q,u}$ represent
the forward and backward difference operator, respectively,
it has essentially the same structure of the fermion kinetic terms of the supersymmetric gauge theory \eqref{eq:Xi}. 
As mentioned above, 
the reason why the fermion doubler is absent in the supersymmetric gauge theory is not the supersymmetry itself but the appearance of the effective Wilson term. 
Therefore, 
although the present model of fermions on the graph is not supersymmetric, 
we can expect that the fermion doubler is absent in our model as well as that of the supersymmetric gauge theory by replacing the incidence matrix $L$ with the deformed one $L_{q,u}$.
%we see that the Dirac operator (\ref{Dirac chiral}) has the same structure as those appearing in supersymmetric gauge theories on the lattice \cite{Sugino:2003yb, Sugino:2004qd,Kikukawa:2008xw,Misumi:2013maa}\footnote{Note that the overall sign of the backward difference operators in our convention differs from the usual one in lattice gauge theory. So the anti-commuting part with $\gamma_5$ becomes anti-symmetric on $L_{q,u}$ and $\Lt_{q,u}$.}.
%In the supersymmetric lattice gauge theory,
%, we can understand the reason why the doublers do not appear in the fermion sector.
% Honestly speaking, since our fermionic model is inspired from supersymmetric quiver gauge theories \cite{Ohta:2014ria,Ohta:2015fpe,Ohta:2020ygi} 
% and supersymmetric gauge theory on the graph \cite{Matsuura:2014kha,Matsuura:2014nga,Kamata:2016xmu,Ohta:2021jze},
% it is actually not surprising that there is a relation between our model and supersymmetric theory.

%The relation between the fermions in the Sugino model
%and the fermions in our model is not so clear at the moment,
%but it is important to point out that the incidence matrix $L$ naturally appears in the supersymmetric transformation \eqref{SUSY transformation} and plays a part of the Dirac operator in the action.
%So we can expect that the fermion doubling problem is avoided in our model as well as the Sugino model
%by replacing the incidence matrix $L$ with the deformed one $L_{q,u}$.

%To support our conclusion, we have plotted the eigenvalue distribution of the Dirac operator on $10\times 10$ square and honeycomb lattice in Fig.~.

%%%
\subsection{Chiral transformation and overlap fermion}

Related to the fermion doubling problem, 
we can introduce a chiral transformation for the fermions on the graph by
\be
\begin{split}
\Psi \rightarrow \Psi' &=e^{i\theta \gamma_5}\Psi \, ,\\
\bar{\Psi} \rightarrow \bar{\Psi}' &=\bar{\Psi}e^{i\theta \gamma_5}\, ,
\end{split}
\label{chiral transformation}
\ee
where $\gamma_5$ is taken to be the Dirac basis
\be
\gamma_5 = \begin{pmatrix}
  I_{n_V} & 0 & 0\\
  0 & 0 & I_{n_E}\\
  0 & I_{n_E} & 0
\end{pmatrix}\,,
\ee
which generates a rotation between the fermions on the edges $\psi$ and $\psit$.

%In the limit of $u\to 0$ and $q\to 1$,
%namely $t\to1$,
%%$\alpha=\sqrt{\frac{q}{1-t^2}}\to\infty$
%%and 
%the Dirac operator $\slashed{D}+{\cal M}$ reduces to 
%%the fermion action (\ref{fermion action})
%%is invariant under the chiral transformation (\ref{chiral transformation}),
%%since $\alpha\to \infty$ in this limit and
%\be
%\slashed{D}+{\cal M} \sim \alpha\begin{pmatrix}
%0 & - L^T &  L^T\\
% L & 0 & 0\\
%-L & 0 & 0  
%\end{pmatrix}\, ,
%\ee
%since $\alpha \to \infty$ in this limit. 
%which anti-commutes with $\gamma_5$.
%Therefore, the fermion action (\ref{fermion action}) is invariant under the chiral transformation (\ref{chiral transformation}) in this limit. 
%However, in the generic parameter region of $u$ and $q$,
%the fermion action (\ref{fermion action}) is not invariant under the chiral transformation. 
%%since not only the explicit mass term ${\cal M}$ violates the chiral symmetry but also $\slashed{D}$ does.
%Thus, at the expense of the chiral symmetry, the fermion doubling problem is expected to be avoided at least in the generic parameter region of $u$ and $q$, similar to the Wilson fermion.

Interestingly, we can show that the massive Dirac operator of our model in the Weyl basis 
\begin{align}
  A \equiv \slashed{D}+{\cal M}
\end{align}
%in short. 
%Related to the property of $\gamma_5$ and the chiral fermions on the Weyl basis, we also find that 
%Interestingly, the massive Dirac operator of our model 
satisfies the $\gamma_5$-hermiticity
\be
\gamma_5 A \gamma_5 = A^\dagger\,. 
\ee
%where we have denoted $A=a(\slashed{D}+{\cal M})$ shortly.
Thus, we can construct the overlap fermion \cite{Neuberger:1997fp,Neuberger:1998wv} as
\be
D_{\rm ov} = \frac{1}{a}\left(1+\frac{A}{\sqrt{A^\dag A}}\right)
= \frac{1}{a}\left(1+\gamma_5{\rm sign}\,(\gamma_5 A)  \right)\, ,
\ee
which satifies the Ginsparg-Wilson relation \cite{Ginsparg:1981bj}
\be
D_{\rm ov}\gamma_5+\gamma_5 D_{\rm ov}= a D_{\rm ov}\gamma_5 D_{\rm ov}\, .
\ee
This is an important result not only for physics as a model on the discrete space-time but also for mathematics.
%It is important problem to consider the overlap fermion on the discrete space-time, but it is also interesting for mathematics.
Since we constructed the Dirac operator to associated with the deformed Laplacian on the graph,
%the eigenvalues of the Dirac operator are closely related to the Laplacian on the graph. %, of course.
%Since the constructed Dirac operator is closely related to the Laplacian on the graph and the graph zeta function,
this result opens up a possibility to study the eigenvalue distributions from the viewpoint of 
%These are related with theory for 
the spectral zeta function or index theorem through the overlap fermions.
Although our interest for these issues is inexhaustible, we will leave it for future work, since they are out of focus on this paper.

%%%
\section{Winding of Cycles and Statistical Mechanics}
\label{sec:winding}

%So far we have introduced two deformed parameters (fugacities) $q$ and $u$,
%which count the length and bumps of the cycles, respectively, 
%as the power series of the Bartholdi zeta function.
In this section, 
in addition to the parameters $q$ and $u$,
we introduce another parameter $r$ to the graph zeta function which counts the winding number of the cycles of a graph. 

%%%
\subsection{Winding number and the graph zeta function}

\begin{figure}[t]
  \begin{center}
  \includegraphics[scale=0.7]{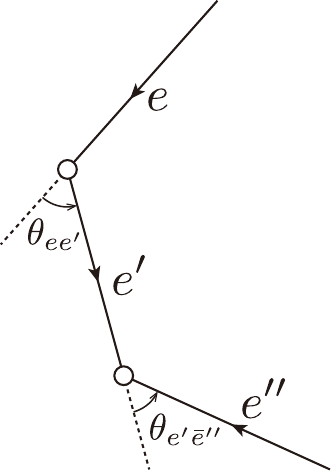}
  \end{center}
  \caption{The relative exterior angles between the directed edges on the plane which measures the rotation of the cycle.}
  \label{exterior angles}
\end{figure}
In order to introduce the winding number, we first draw the graph on the two-dimensional Euclidean plane and define the exterior oriented angle $\theta_{\bse\bse'}$ ($-\pi\leq\theta_{\bse\bse'}\leq\pi$) between the edges $\bse$ and $\bse'$ by regarding the directed edges as vectors on the plane (see Fig.~\ref{exterior angles}). 
By definition, it satisfies $\theta_{\bse\bse'}=-\theta_{\bse'\bse}$.
Furthermore, we assign 
%the exterior oriented angle to an edge and its inverse 
the angle between $e\in E$ and its inverse $\bar{e}$
as $\theta_{e\bar{e}}=\pi$ and $\theta_{\bar{e}e}=-\pi$.
% for $e\in E$. 
Then, 
%we can define 
the winding number $w(C)$ 
of a cycle $C=\bse_{1}\bse_{2}\cdots \bse_{k}$ is defined as
\begin{align}
  w(C) \equiv \frac{1}{2\pi} \sum_{i=1}^k \theta_{\bse_{i}\bse_{{i+1}}}\, ,
\end{align}
with the convention $\bse_{k+1}=\bse_{1}$.
The winding number takes on 
a positive or negative integer value depending on whether the cycle rotates counterclockwise or clockwise.

%The winding number of the cycle $C$ is represented by $w(C)$, which takes positive
%or negative integers depending on the counterclockwise or clockwise rotation.
%To introduce the winding number into the graph zeta function,
%As a preparation to introduce the winding number into the cycles of the graph, 
%we assign an exterior oriented angle $\theta_{\bse\bse'}$ between the edges $\bse$ and $\bse'$, 
%angle between two directed edges %an angle on a plane
%and denote the exterior oriented angle 
%as $\theta_{\bse\bse'}$,
%which satisfies $\theta_{\bse\bse'}=-\theta_{\bse'\bse}$. (See Fig.~\ref{exterior angles}.)
%Using these angles, 
Correspondingly, 
we define the weighted edge adjacency matrix $W(r)$ and the weighted bump matrix $J(r)$ as 
\be
\begin{split}
  W_{\bse \bse'}(r) &= \begin{cases}
  r^{\theta_{\bse\bse'}/2\pi} &\text{if $t(\bse')=s(\bse)$ and $\bse'\neq \bar{\bse}$}\\
  0 & \text{others}
  \end{cases}\, ,\\
  J_{\bse \bse'}(r) &= \begin{cases}
    r^{1/2} &\text{if $\bse\in E$ and $\bse' = \bar{\bse}$}\\
    r^{-1/2} &\text{if $\bse'\in E$ and $\bse' = \bar{\bse}$}\\
    0 & \text{others}
    \end{cases}\, ,
    \label{eq:WJ with r}
  \end{split}
\ee
%where $\theta_{\bse\bse'}\equiv \theta_{\bse'}-\theta_{\bse}$.
%Note here that we need to define suitably the sign of the angle
%depending on the order of the directions,
%including the part of the bumps where $\theta_{\bse\bse'}=\pm \pi$.
%the non-zero elements in $J$ takes the value of $r^{\pm 1/2}$ only,
%since the matrix $J$ corresponds to the bump part of the cycle.
and define a graph zeta function with windings %in the Hashimoto expression defined by $W(r)$ and $J(r)$ 
as
\be
\tilde{\zeta}_\Gamma(q,u,r)^{-1}
\equiv
\det \left(I_{2n_E}-q B_u(r)\right)\, ,
\label{eq:graph zeta with winding}
\ee
where $B_u(r)=W(r)-uJ(r)$.
By using the general formula for the weighted Bartholdi zeta function or Amitsur's theorem \cite{amitsur1980characteristic, 
reutenauer1987formula}, 
it is clear that \eqref{eq:graph zeta with winding} can be expressed by the Euler product
\be
\tilde{\zeta}_\Gamma(q,u,r)^{-1}
= \prod_{[C]:\substack{\text{primitive}}}(1-r^{w(C)}u^{b(C)}q^{\ell(C)})\, .
\ee
%then we can show that it has the following Euler product representation
%\be
%\tilde{\zeta}_\Gamma(q,u,r)^{-1}
%= \prod_{[C]:\text{primitive}}(1-\alpha_C(r)u^{b(C)}q^{\ell(C)})\, ,
%\ee
%where
%\be
%\alpha_C(r) = r^{\theta_{\bse_1\bse_2}/2\pi}
%\times
%r^{\theta_{\bse_2\bse_3}/2\pi}
%\times
%\cdots
%\times
%r^{\theta_{\bse_k\bse_1}/2\pi}\,,
%\ee
%for a cycle $C=\bse_1\bse_2\cdots \bse_k$,
%by using the general formula for the weighted Bartholdi zeta function or Amitsur's theorem.
%By definition, it is clear that 
%%$\alpha_C(r)=r^{w(C)}$
%since the sum of the exterior angles around the cycle
%counts the number of the rotations (windings) on the plane.
%So we finally obtain more generic graph zeta function
%with the windings
%\be
%\tilde{\zeta}_\Gamma(q,u,r)^{-1}
%= \prod_{[C]:\substack{\text{primitive}\\\text{reduced\ \,}}}(1-r^{w(C)}u^{b(C)}q^{\ell(C)})\, .
%\ee
Furthermore, 
by repeating the same argument in Sec.~\ref{sec:fermionic cycles}, 
%Using the notion of the fermionic cycles and the cycle M\"obius function, the above 
we can rewrite it in the power series up to the finite order $2n_E$ in $q$ as
\be
\begin{split}
\tilde{\zeta}_\Gamma(q,u,r)^{-1}
&=1+\sum_{[C_P]}\mu(C_P)
r^{w(C_P)}u^{b(C_P)}q^{\ell(C_P)}\, .
\end{split}
\ee
%as same as the original Bartholdi zeta function. 

Let us see examples of the graph zeta functions with windings.
By considering the $C_3$ graph as an equilateral triangle on the plane
as depicted in Fig.~\ref{cycle graph C3},
the matrices $W(r)$ and $J(r)$ are given by
\be
%\begin{split}
\mathscale{0.8}{
W_{\bse \bse'}(r) = \begin{pmatrix}
0 & r^{1/3} & 0 & 0 & 0 & 0\\
0 & 0 & r^{1/3} & 0 & 0 & 0\\
r^{1/3} & 0 & 0 & 0 & 0 & 0\\
0 & 0 & 0 & 0 & 0 & r^{-1/3}\\
0 & 0 & 0 & r^{-1/3} & 0 & 0\\
0 & 0 & 0 & 0 & r^{-1/3} & 0
\end{pmatrix}\,,\quad 
J_{\bse \bse'}(r) = \begin{pmatrix}
  0 & 0 & 0 & r^{1/2} & 0 & 0\\
  0 & 0 & 0 & 0 & r^{1/2} & 0\\
  0 & 0 & 0 & 0 & 0 & r^{1/2}\\
  r^{-1/2} & 0 & 0 & 0 & 0 & 0\\
  0 & r^{-1/2} & 0 & 0 & 0 & 0\\
  0 & 0 & r^{-1/2} & 0 & 0 & 0
  \end{pmatrix}\, ,
}
%\end{split}
\ee
respectively. 
Then, we obtain
\be
\begin{split}
%\tilde{\zeta}_{C_3}(q,u,r)^{-1}&=
%1-3 u^2 q^2 -(r+r^{-1})q^3-(3u^2-3u^4) q^4+(1-3u^2+3u^4-u^6)q^6\, .
\tilde{\zeta}_{C_3}(q,u,r)&=
\left(1-3 u^2 q^2 -(r+r^{-1})q^3-(3u^2-3u^4) q^4+(1-3u^2+3u^4-u^6)q^6\right)^{-1} \\ 
&= 
1 + 3u^2q^2 + \left(r+r^{-1}\right)q^3 + (6u^4+3u^2)q^4+6u^2(r+r^{-1})q^5+{\cal O}(q^6)\,.
\end{split}
\ee
From the coefficient of $q^3$ in the second line, 
the cycles of length $3$ have the winding number $1$ and $-1$ as expected, 
%where we see that two different cycles with the winding number $w(C)=1$ and $w(C)=-1$
%of length 3 ($q^3$ term) in the cycle graph $C_3$.
%the cycles including the bumps does not include non-reduced cycle and the winding number, there $r$ does not appear in the term with $u$.
and 
the cycles that collapse to a single point by reducing bumps does not contain $r$ 
as seen in the terms of $q^2$ and $q^4$. 

Similarly, for the double triangle graph depicted in Fig.~\ref{double triangle graph}, 
we find, at $u=0$,
\be
\begin{split}
\tilde{\zeta}_{\rm DT}(q,r)^{-1}&=
1-2 \left(r+r^{-1}\right)q^3
-\left(r+r^{-1}\right)q^4\\
&\qquad+\left(r^2+2+r^{-2}\right)q^6
+4 q^7+q^8
- (r+2+r^{-1})q^{10}\,,
\end{split}
\ee
which again well explains the winding numbers of the cycles in the double triangle graph.

%%%
\subsection{Gauge invariance}

We here point out that the assignment of the angle $\theta_{\bse\bse'}$ is not unique to define the same graph zeta function \eqref{eq:graph zeta with winding}. 
To see it, we assign weight $\alpha_v$ ($\alpha_v \in \R$) at each vertex $v\in V$ and define a matrix $R_\alpha$ of size $2n_E$ as 
\begin{align}
  R_\alpha \equiv \diag\left(r^{-\alpha_{s(\bse_1)}/2\pi},r^{-\alpha_{s(\bse_2)}/2\pi},
  \cdots,
  r^{-\alpha_{s(\bse_{2n_E})}/2\pi}\right)\,.
  \label{eq:gauge trans matrix}
\end{align}
Then, we modify the weighted edge adjacency matrix $W(r)$ as
\begin{align}
  W_\alpha(r) \equiv R_\alpha W(r) R_\alpha^{-1}\,,
  \label{eq:deformation of W}
\end{align}
which shifts the weight $\theta_{\bse\bse'}$ as 
\be
\theta_{\bse\bse'} \to \theta_{\bse\bse'} +\alpha_{t(\bse)}-\alpha_{s(\bse)}\,.
%\begin{split}
%  W_{\bse \bse'}(r) &= \begin{cases}
%  r^{\frac{\theta_{\bse\bse'}+\alpha_{t(\bse)}-\alpha_{s(\bse)}}{2\pi}} &\text{if $t(\bse')=s(\bse)$ and $\bse'\neq \bar{\bse}$}\\
%  0 & \text{others}
%  \end{cases}\,.
%\end{split}
\label{eq:shift}
\ee
Apparently, the graph zeta function with windings $\tilde{\zeta}_\Gamma(q,u,r)$ is invariant under this transformation since it is defined by the determinant \eqref{eq:graph zeta with winding} and the matrix $J$ is invariant under this transformation.
This is simply because the deformation \eqref{eq:deformation of W} does not change the winding numbers of the cycles. 

This is nothing but the gauge invariance of the graph zeta function under the local transformation of the weights at the vertices by the group $\R_+$.
We can understand it by looking at the gauge transformation of the effective action of the FKM model defined in \cite{Matsuura:2023ova}, 
which is a kind of lattice gauge theory on the graph as mentioned in Introduction.
In the case of the FKM model, 
we put unitary matrices of size $N_c$ on the edges as link variables
and 
the effective action is the unitary matrix weighted Bartholdi zeta function \cite{Matsuura:2022ner,Matsuura:2022dzl}. 
As same as the usual lattice gauge theory, 
the theory is invariant under the $U(N_c)$ gauge transformation which locally rotates the variables on the vertices. 
By setting $N_c=1$, 
the link variable on the edge $\bse$ takes the value of $U(1)$ as $e^{i\theta_{\bse}}$ 
and the gauge group reudces to $U(1)$. 
Furtherore, by rotating $\theta_{\bse}$ as $\theta_\bse \to i \theta_\bse$, 
the link variable becomes the positive real number $e^{-\theta_\bse}$  
and the gauge group of the FKM model reduces to the non-compact Abelian group $\R_+$. 
Although the effective action of the modified FKM model does not coincide to the graph zeta function with windings, 
the gauge transformation of the weighted edge adjacency matrix is exactly equal to \eqref{eq:gauge trans matrix}. 
This shows that the transformation \eqref{eq:shift} can be regarded as a gauge transformation.

%%%
\subsection{Connection to the Ising model}

Once we include the winding number $r$ into the graph zeta function,
we encounter the interesting relationship between the zeta function and the partition function of the statistical mechanics.
Historically, it is first discovered by Kac and Ward that the partition function of the Ising model on the two-dimensional square lattice is expressed in terms of the edge adjacency matrix 
with signatures, which is called Kac-Ward matrix \cite{PhysRev.88.1332}.
The Kac-Ward matrix is equivalent to the deformed edge adjacency matrix \eqref{eq:WJ with r} at $r=-1$.
So the partition function of the two-dimensional Ising model on the generic graph,
which is called the random bond Ising model (RBIM), can be obtained at a special value of the graph zeta function
\be
\begin{split}
\tilde{\zeta}_\Gamma(q,u{=}0,r{=}{-}1)^{-1}
&=\prod_{[C]:\substack{\text{primitive}\\\text{reduced\ \,}}}\left(1-(-1)^{w(C)}q^{\ell(C)}\right)\\
&=2^{-2n_V}(1-q^2)^{n_E}\bigl(Z_\Gamma^{\text{Ising}}\bigr)^2\, ,
\end{split}
\label{Ising/Zeta correspondence}
\ee
where $q=\tanh(\beta J)$, and $\beta$ and $J$ are the inverse temperature and coupling of the Ising model, respectively.

This means that, 
%for our model of fermions on the graph, we can obtain a fermion system whose 
the partition function of our model is equivalent to that of the RBIM by considering an appropriate charge corresponding to the number of windings (holonomy or magnetic flux associated with each cycle)
\be
Z_F(q,u{=}0,r{=}{-}1)
={\cal N}\beta^{n_V+2n_E}2^{-2n_V}(1-q^2)^{n_E}\bigl(Z_\Gamma^{\text{Ising}}\bigr)^2
\, .
\ee
It is well known fact that there is a correspondence between the Ising model and free fermion system, but this gives a new perspective on this relationship for the RBIM on the general graph.

Let us see concrete examples of the correspondence between the graph zeta function and the partition function of the RBIM.

%For the cycle graph $C_N$, it is a simple excercise of the Ising model on the one-dimensional chain.
The Ising model on the one-dimensional chain, namely, on the cycle graph $C_N$ is a simple exercise. 
The Hamiltonian is given by
\be
H = -J \sum_{i=1}^N \sigma_i\sigma_{i+1},
\ee
with the periodic boundary condition $\sigma_{L+1}=\sigma_1$. 
Then, the partition function can be written as 
\be
Z_{C_N}^{\text{Ising}}
=\sum_{\sigma_1=\pm 1}
\cdots
\sum_{\sigma_N=\pm 1}e^{-\beta H}=\Tr T^N\,,
\ee
with the transfer matrix, 
\be
T = \begin{pmatrix}
e^{\beta J} & e^{-\beta J}\\
e^{-\beta J} & e^{\beta J}
\end{pmatrix}
\, , 
\ee
which reduces to 
\be
Z_{C_N}^{\text{Ising}}=\lambda_+^N+\lambda_-^N \,,
\ee
where $\lambda_{\pm}=e^{\beta J}\pm e^{-\beta J}$ are the eigenvalues of $T$. 
By setting $q=\tanh(\beta J)$,
we see 
\be
\tilde{\zeta}_{C_N}(q,u{=}0,r{=}{-}1)^{-1}
=(1+q^N)^2
=2^{-2N}(1-q^2)^N \bigl(Z_{C_N}^{\text{Ising}}\bigr)^2\, ,
\ee
which agrees with (\ref{Ising/Zeta correspondence}).

Another example is the double triangle graph depicted in Fig.~\ref{double triangle graph}.
The Hamiltonian of the Ising model on the double triangle graph is written by using the adjacency matrix $A$ as
\be
H= -J\sum_{v,v'\in V} A_{vv'}\sigma_v\sigma_{v'},
\ee
since the adjacency matrix expressed the nearest neighbor interaction on the graph.
On the double triangle graph, 
%there are four vertices, so 
there exists $2^4=16$ spin configurations since it has four vertices. 
By adding up the contribution from each configuration, the partition function can be evaluated as 
%If we evaluate the partition function by counting possible configurations, we find
\be
\begin{split}
Z_{\rm DT}^{\text{Ising}}
&=\sum_{\{\sigma_v=\pm 1|v\in V\}}e^{-\beta H}\\
&=2\left(
  e^{5\beta J}+2e^{\beta J}+4e^{-\beta J}+e^{-3\beta J}
\right)\, . 
\end{split}
\ee
On the other hand, the graph zeta function of the double triangle graph with windings at $r=-1$ is given by
\be
\begin{split}
\tilde{\zeta}_{\rm DT}(q,u=0,r=-1)^{-1}
&=(1+q)^2 \left(1-q+q^2+q^3\right)^2\\
&=\frac{16e^{4\beta J}
\left(1
+3e^{2\beta J}-e^{4\beta J}+e^{6\beta J}
\right)^2}{\left(1+e^{2\beta J}\right)^8}
\, ,
\end{split}
\ee
which agrees with $2^{-8}(1-q^2)^5\bigl(Z_{\rm DT}^{\text{Ising}}\bigr)^2$.

Combining the discussion involving the windings here with the $L$-function of the grid graph discussed in the previous section reveals quite interesting properties of the fermion system on the graph.
Let us recall that the poles of the graph zeta function are the zeros of the partition function of our fermionic model.
According to the Lee-Yang circle theorem \cite{PhysRev.87.410},
the zeros of the partition function of the statistical models are distributed in the complex plane of a parameter (fugacity) as a circle in the thermodynamic limit and the phase transition point is located on the real axis separated by the circle.
In our fermionic model, the parameter $q$ is the parameter itself in the Lee-Yang theorem
and the poles of the graph zeta function in the complex $q$-plane are expected to relate to the phase transition points. 

\begin{figure}[ht]
  \begin{center}
  \subcaptionbox{$100\times 100$ square lattice}[.45\textwidth]{
  \includegraphics[scale=0.7]{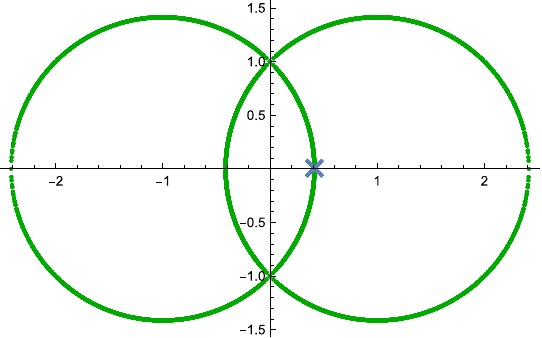}
  }
  \subcaptionbox{$100\times 100$ honeycomb lattice}[.45\textwidth]{
  \includegraphics[scale=0.7]{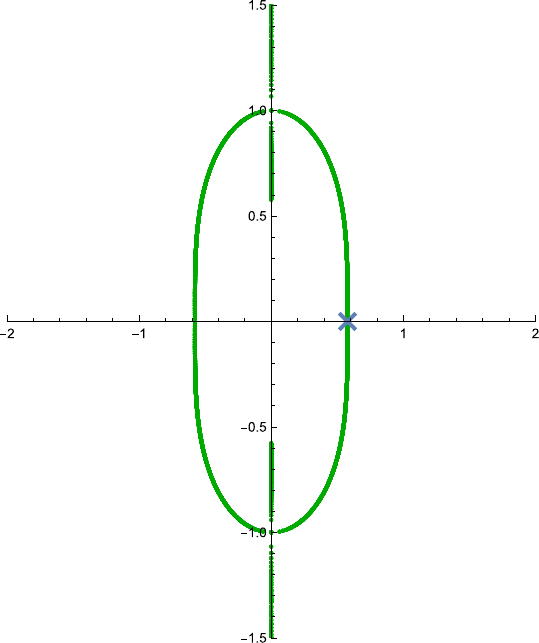}
  }
  \end{center}
  \caption{The poles of the Ihara zeta function with windings (zeros of the partition function of the fermionic model)
   at $r=-1$, which is equivalent to the zeros of the two-dimensional Ising model, for the $100\times 100$ square lattice (a) and the $100\times 100$ honeycomb lattice (b) in the complex $q$-plane.
  The cross markers represent the phase transition points in $q$, which are $q_c=0.414214\cdots$ for the square lattice and $q_c=0.57735\cdots$ for the honeycomb lattice, respectively.}
  \label{Lee-Yang circle}
\end{figure}

By repeating the construction of the Bartholdi zeta function on the grid graph by using the Artin-Ihara $L$-function, 
we can construct the graph zeta function with windings on the grid graph. 
For example, the Ihara zeta function of the square lattice with windings is given by
\be
\zeta_{\rm SQ}(q,r)^{-1}
=\prod_{m_1=0}^{N-1}\prod_{m_2=0}^{M-1}
\left\{(1-q^2)\left(
1+3q^2
-q \hat{A}_{\rm SQ}(\vec{m})\right)
-(r^{1/2}-r^{-1/2})^2q^4
\right\}\, ,
\ee
%and investigate the distribution of the poles in the complex $q$-plane.
and that of the honeycomb lattice is given by
\be
\begin{split}
  \zeta_{\rm HC}(q,r)^{-1}
  &=\prod_{m_1=0}^{N-1}
  \prod_{m_2=0}^{M-1}
 \left\{ (1-q^2)\det\left((1+2q^2)I_2-q\hat{A}_{\rm HC}(\vec{m})\right)-(r^{1/2}-r^{-1/2})^2q^6\right\}\, .
\end{split}  
\ee
The distribution of the poles of the graph zeta function drastically changes by including the windings from that of the original Bartholdi zeta function shown in Fig.~\ref{zeros of the grids}.
The results of the distribution of the poles of the Ihara zeta function (zeros of the partition function of the fermionic model) with windings at $r=-1$ for the $100\times 100$ square lattice and the $100\times 100$ honeycomb lattice in the complex $q$-plane are shown in Fig.~\ref{Lee-Yang circle}.

The Ising model on the square lattice has been exactly solved by Onsager \cite{PhysRev.65.117} and the phase transition point is given by the solution of
\be
\sinh(2\beta_c J) = 1 \, ,
\ee
which corresponds to $q_c=\tanh(\beta J)=0.414214\cdots$.
We see the circle in the complex $q$-plane acrosses the real axis at $q_c$ in Fig.~\ref{Lee-Yang circle} (a).
For the honeycomb lattice, the phase transition point is given by the solution of \cite{HOUTAPPEL1950425}
\be
\sinh(2\beta_c J) = \sqrt{3}\, , 
\ee
which corresponds to $q_c=\tanh(\beta_c J)=0.57735\cdots$.
The phase transition point again appears at the crossing point of the circle with the real axis in Fig.~\ref{Lee-Yang circle} (b).

\section{Covering Graph, $L$-function and Gauge Theory}
\label{sec:gauge}

%In this section, %we would like to 
%we generalize the fermion representation of the graph zeta function to the graph with the gauge field. 

%In \cite{Ohta:2021jze,Matsuura:2022dzl,Matsuura:2022ner,Matsuura:2023ova,Matsuura:2024gdu},
%In \cite{Matsuura:2022dzl,Matsuura:2022ner,Matsuura:2023ova,Matsuura:2024gdu,Matsuura:2024rcv},
As mentioned in Introduction, 
gauge theories called the generalized Kazakov-Migdal models are constructed on the graph as a kind of lattice gauge theory,
where 
the scalar field is defined on the vertices and the gauge field (link variables) are defined on the edges of the graph. %respectively. 
%\cite{Matsuura:2022dzl,Matsuura:2022ner,Matsuura:2023ova,Matsuura:2024gdu,Matsuura:2024rcv}.
%The scalar field on $V$ is the element of the representation space ${\cal H}_R$,
%and the gauge field $E$ is represented by unitary matrices as link variables.
The scalar field of the model belongs to the adjoint representation in \cite{Matsuura:2022dzl,Matsuura:2022ner} or to the fundamental representation in \cite{Matsuura:2023ova,Matsuura:2024gdu}.
%The partition function of the generalized Kazakov-Migdal model is expressed as the path integral over the Artin-Ihara $L$-function emerging from the derived covering graph $\Gamma^G$.
In both cases,
the partition functions %generalized Kazakov-Migdal model on the graph 
are given by the path integral over the unitary matrix weighed graph zeta function where the unitary matrices are acting on the adjoint or fundamental representation. 

%A graph whose group elements are placed on the edges is 
%which derives 

\begin{figure}[ht]
\begin{center}
\includegraphics[scale=0.8]{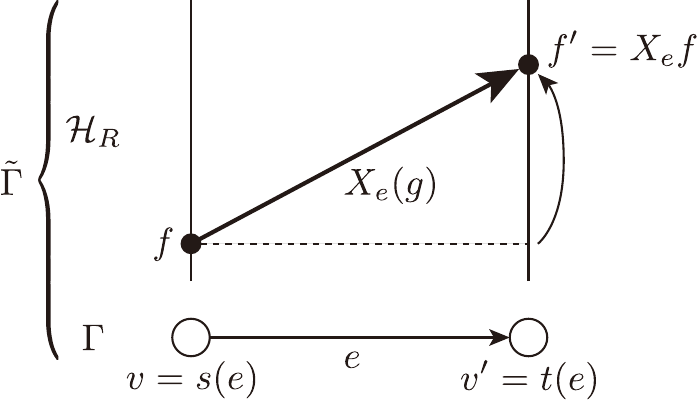}
\end{center}
\caption{An image of the derived graph $\tilde\Gamma$ over an edge of the base graph $\Gamma$ with the voltage assignment of the group $G$.
A representation of the group assigned on the edge $X_e(g)$ induces an action on the representation space ${\cal H}_R$ at the target vertex.}
\label{derived covering graph}
\end{figure}

In Sec.~\ref{sec:covering graph}, 
we have defined a covering graph by assigning a group element on each edge of the base graph, which is known as a voltage graph in graph theory.
%which is known as a voltage graph in the graph theory. 
Instead,  
we can assign a $d_R$-dimensional irreducible representation $X_e\in GL(d_R,\mathbb{C})$ of the group element $g_e\in G$ on the edges.
Correspondingly,
let ${\cal H}_R$ be a $d_R$-dimensional representation space.
%The voltage assignment of vertices on 
In this setting, 
the vertices of the derived graph $\tilde\Gamma$ 
is given by pairs $(v,f)\in V\times {\cal H}_R$, where $v$ is a vertex of the base graph $\Gamma$ and $f\in {\cal H}_R$.
%The voltage assignment of the 
The edges of $\tilde\Gamma$ 
is given by the pairs of the neighborhood vertices $\langle (v,f), (v',X_e f)\rangle$
for each edge $e=\langle v,v'\rangle\in E$ of the base graph $\Gamma$. 
%and $X_e:E\to G$ is a map from the edge $e\in E$
%to the representation of $G$.
The voltage assignment gives the fiber bundle structure on the derived graph $\tilde\Gamma$
and there is a natural projection map $\pi:\tilde\Gamma\to \Gamma$.
We have depicted an image of the covering graph in Fig.~\ref{derived covering graph}.

%If we consider 
%it reduces to 
In this terminology, 
the unitary matrix weighted Bartholdi zeta function is nothing but the Artin-Ihara $L$-function on the base graph $\Gamma$
%By definition, 
%the Artin-Ihara $L$-function on the base graph $\Gamma$
%is equal to 
%the graph zeta function of the derived covering graph $\tilde\Gamma$
\be
%\begin{split}
L_\Gamma(q,u;X)
\equiv
\prod_{[C]:\text{primitive}}
\det\left(1-X_C\, u^{b(C)}q^{\ell(C)}\right)^{-1} 
\, ,
%\\ 
%&=
%\zeta_{\Gamma^{G,R}}(q,u)
%\end{split}
\label{Artin-Ihara L-function}
\ee
where $X_C$ is the ordered product of $X_{\bse}$ around the cycle $C$ with the understanding that $X_{\bar{e}}=X_{e}^{-1}$.
Therefore, 
the generalized Kazakov-Migdal model on the graph is also regarded as a theory of the derived graph $\tilde\Gamma$ by the voltage assignment of the unitary group $G=U(N)$.

%In the grid graph (the derived covering graph of the finite group), it is natural to take the regular representation, namely $X_e(g)=g$.
%According to the Peter-Weyl theorem, the regular representation of the finite group is decomposed into the irreducible representations of the group as
%\be
%X_{\rm reg}=\bigoplus_{R:{\rm irr.rep.}}d_R X_R\, ,
%\ee
%where $d_R$ is the dimension of the representation $R$.

The purpose of this section is to express the {\it inverse} of the Artin-Ihara $L$-function (\ref{Artin-Ihara L-function}) by the partition function of a theory of fermions on the graph. 
Let us first define the deformed adjacency matrix 
\begin{align}
  \left(A_X\right)_{vv'} 
  = \sum_{e\in E} \left( X_e \delta_{v,s(e)}\delta_{v',t(e)} + X_e^{-1} \delta_{v,t(e)}\delta_{v',s(e)} \right)\,, 
\end{align}
and the deformed covariant graph Laplacian
\be
\Delta_{q,u}(X)\equiv
I_{d_Rn_V}-qA_X+q^2(1-u)I_{d_R}\otimes(D-(1-u)I_{n_V})\, .
\ee
%where $D$ is the tensor product of the degree matrix of the base graph $\Gamma$ and $I_{d_R}$.
Then, by using Ihara's theorem, 
the inverse of the Artin-Ihara $L$-function (\ref{Artin-Ihara L-function})
is expressed as 
\begin{align}
  L_\Gamma(q,u;X)^{-1} 
  =
  (1-q^2(1-u)^2)^{d_R(n_E-n_V)}\det\Delta_{q,u}(X)\,.
\end{align}

In order to express the Artin-Ihara $L$-function by the partition function of the fermions on the graph,
we define the deformed incidence matrices
(gauge covariant difference operator)
on the derived graph $\tilde\Gamma$
\be
L_{q,u}(X)\equiv T_X-t S\,,
\qquad
\Lt_{q,u}(X)\equiv S-tT_X \,,
\ee
where $t=q(1-u)$ and
\be
{(T_X)^e}_v=
\begin{cases}
X_e & \text{if $v=t(e)$}\\
0 & \text{others}
\end{cases}\,.
\ee
These incidence matrices are rectangular matrices of size 
$d_R n_E   \times  d_R n_V $ and are analogous to the covariant derivative in the gauge theory on a differentiable manifold.
%Requiring that the transposition of $T$ consists of the inverse\footnote{
%  For the unitary representation, the transpose is simply replaced with the Hermitian conjugate.}
%of $X_e$, the adjacency matrix and degree matrix are expressed as follows:
%In the same way, we define 
It is easy to show that the deformed 
adjacency matrix and degree matrix are expressed by
\be
A_X \equiv S^TT_X+T_{X^{-1}}^TS\, ,
\qquad
D\equiv T_{X^{-1}}^TT_X+S^TS=T^TT+S^TS \, .
\ee
%Note that the degree matrix $D$ still does not depend on $X_e$.

Then, let us consider a model with the action 
\begin{align}
  S \equiv \Tr_R\left\{ \bar\Psi \left(\slashed{D}(X)+{\cal M}\right) \Psi \right\}\,,
\end{align}
where $\Psi$ and $\bar\Psi$ are extensions of the fermions \eqref{eq:fermions on the graph} whose elements take values in the representation ${\cal H}_R$ 
and 
the Dirac operator is defined by 
\be
\slashed{D}(X)+{\cal M} \equiv
\begin{pmatrix}
I_{d_R n_V} & \alpha \Lt_{q,u}^T(X^{-1}) & \alpha L_{q,u}^T(X^{-1})\\
\alpha L_{q,u}(X) & I_{d_R n_E} & -tI_{d_R n_E} \\
\alpha \Lt_{q,u}(X) & -tI_{d_R n_E} & I_{d_R n_E}
\end{pmatrix}\, .
\ee
By repeating the same argument of the matrix decomposition \eqref{eq:matrix decomposition}, we see that the partition function is evaluated as 
\be
Z = \int d\Psi d\bar\Psi e^{-S}
= 
\det \left(\slashed{D}(X)+{\cal M}\right)
=(1-q^2(1-u)^2)^{d_R(n_E-n_V)}\det\Delta_{q,u}(X)\, ,
\ee
which reproduces the inverse of the Artin-Ihara $L$-function. 

We can also repeat the same argument to obtain the Hashimoto expression of the fermion determinant
\be
\det \left(\slashed{D}(X)+{\cal M}\right)
=\det \left(I_{2d_R n_E}-q B_u(X)\right)\, ,
\ee
where $B_u(X)=W(X)-uJ$ and $W(X)$ is a matrix-weighted edge adjacency matrix
\be
W(X)_{\bse \bse'} = \begin{cases}
X_{\bse} &\text{if $t(\bse')=s(\bse)$ and $\bse'\neq \bar{\bse}$}\\
0 & \text{others}
\end{cases}\, ,
\ee
with $X_{\bar{e}}=X_{e}^{-1}$.
Using this expression, we find that the partition function is expressed by a $2d_R n_E$-th order polynomial in $q$ as
\be
L_\Gamma(q,u;X)^{-1}
=1+\sum_{[\tilde{C}_P]}\mu(\tilde{C}_P)
{\cal W}_{\tilde{C}_P}(X)
u^{b(\tilde{C}_P)}q^{\ell(\tilde{C}_P)}\, ,
\ee
where $\ell(\tilde{C}_P)\leq 2d_R n_E$,
$\tilde{C}_P$ stands for a fermionic cycle on the whole derived graph $\tilde\Gamma$,
and ${\cal W}_{\tilde{C}_P}(X)$ is a gauge invariant Wilson loop operator
associated with the fermionic cycle $\tilde{C}_P$.
Using the expression of the Artin-Ihara $L$-function (\ref{Artin-Ihara L-function}),
${\cal W}_{\tilde{C}_P}(X)$ can be expressed
in terms of a multi-trace operator (character)
of the representation matrix $X_R$ and partitions (Young tableaux), but
the details of the fermionic cycle for non-Abelian gauge theory are not the focus of this paper and will be reported in a different context in the near future.

Finally, we would like to comment on the wider applications of this covering graph in gauge theories.
As discussed in Sec.~\ref{sec:grid graph}, 
the discretized space-time used in the conventional lattice gauge theory can be regarded as the grid graph arising from a special kind of the covering graph of an Abelian group.  %which is regarded as a ``seed'' to construct the space-time structure. 
%the graph zeta function of the grid graph is expressed in terms of a product of the Artin-Ihara $L$-function. 
%Recalling that the grid graph discussed in Sec.~\ref{sec:grid graph} is a special case of the derived covering graph of the finite group,
Combining this Abelian group with the gauge group as a direct product,
the space-time structure and gauge symmetry should be encoded into a huge group and its representation.
This implies a similar philosophy to the reduced matrix model or deconstruction where the space-time structure emerges from the large size of matrix in a suitable representation.
It is also interesting to consider the emergence of the space-time and gauge theory from the graph zeta and $L$-function on the covering graph.

%Thus, by considering the derived covering graphs and associated $L$-function, we were able to construct fermions coupled with the gauge field on the graph, but we can also discuss the grid graphs, as explained so far, from a unified viewpoint as gauge theory on the graph.

\section{Conclusion and Discussions}
\label{sec:conclusion}

In this paper, we have constructed a model of fermions on the graph associated with the graph zeta function.
Our model has various significant properties, such as the generating function of the fermionic cycles, the absence of the fermion species doublers, 
the construction of the overlap fermion emerging from the $\gamma_5$-hermiticity, 
the correspondence to the statistical model (Ising model) on the graph,
and the relationship between the gauge theory and covering graph. 
%More research needs to be done on any of these properties.

Further development can be expected for any of these properties. 
The model of fermions on graphs is expected to have applications not only to lattice gauge theory on graphs, but also to a variety of physics, including condensed matter physics, quantum information theory, and quantum gravity.
%We expect that our model of the fermions on the graph will be useful for various applications in physics, including not only the lattice gauge theory on the graph but also the condensed matter physics, the quantum information theory, and the quantum gravity.
For example, by developing the construction of the domain wall fermion or the index theorem for the Dirac operator on the graph, we can discuss the topological properties of the fermions on the graph and expect to apply them to the topological insulators or the topological superconductors.
In fact, the zeros of the partition function, namely the inverse of the graph zeta function, implies the appearance of the zero mode (massless mode) of the Dirac operator.
%It is a very interesting question to investigate the behavior of the zero modes and spectrum of the Dirac operator depending on the parameters of the graph zeta function, in connection with the topology of the graph.
In connection with the topology of the graph, it is a very interesting problem to study the zero modes and spectral behavior of the Dirac operator depending on the parameters of the graph zeta function.

Finally, we also would like to point out the relation to the supersymmetric gauge theory on the graph \cite{Matsuura:2014kha,Matsuura:2014nga,Kamata:2016xmu,Ohta:2021jze}.
In series of our accomplishments \cite{Matsuura:2022dzl,Matsuura:2022ner,Matsuura:2023ova,Matsuura:2024gdu}, we have proposed the bosonic model whose partition function is expressed in terms of the graph zeta function.
On the other hand, the fermionic model on the graph constructed in this paper gives the inverse of the graph zeta function as the partition function.
By combining the bosonic and fermionic models on the graph, it is possible to impose a supersymmetry (or a BRST symmetry) on the graph, 
which is expected to be useful for the study of the supersymmetric gauge theory on the lattice.
This supersymmetric gauge theory on the graph also should be related to the supersymmetric quiver gauge theories
\cite{Ohta:2014ria,Ohta:2015fpe,Ohta:2020ygi}.
These correspondences lead further understandings of the counting of the gauge invariant (BPS) operators and the superconformal index from the viewpoint of the graph zeta functions.
We will report on these topics in the near future.

\section*{Acknowledgments}
The authors would like to thank
D.~Kadoh,
O.~Morikawa
and
K.~Okunishi
for useful discussions and comments.
This work is supported in part
by Grant-in-Aid for Scientific Research (KAKENHI) (C), Grant Number 23K03423 (K.~O.).

\bibliographystyle{ptephy.bst}
\bibliography{refs}

\end{document}